\documentclass[twocolumn,10pt,journal,a4paper]{IEEEtran}
\usepackage{bbm}
\newcommand{\bm}[1]{\mbox{\boldmath{$#1$}}}
\usepackage[latin1]{inputenc}
\usepackage{amssymb}
\usepackage{latexsym}
\usepackage{mathrsfs}
\usepackage{amsfonts}
\usepackage{amsbsy}
\usepackage{amsmath}
\usepackage{times}
\usepackage{epsfig,epsf,times,amssymb,amsmath}
\usepackage{amsfonts}
\input epsf

\title{\Huge Frequency-Domain Group-based Shrinkage Estimators for UWB Systems}%for SC-FDE in DS-UWB Systems}
\author{Sheng Li, {\it IEEE Member}, Rodrigo C. de Lamare, {\it IEEE Senior Member}, and Martin Haardt, {\it IEEE Senior Member}
\thanks{Copyright (c) 2013
IEEE. Personal use of this material is permitted. However, permission
to use this material for any other purposes must be obtained from the
IEEE by sending a request to pubs-permissions@ieee.org.

Sheng Li is with the College of Information Engineering, Zhejiang University of Technology, China (e-mail: shengli@zjut.edu.cn);
Rodrigo C. de Lamare is with the Communications Research Group, Department of
Electronics, University of York, UK (e-mail: rcdl500@ohm.york.ac.uk);
Martin Haardt is with the Communications Research Laboratory, Ilmenau University of Technology, Germany (e-mail: martin.haardt@tu-ilmenau.de)

This work has been supported by Zhejiang Key Laboratory for Signal Processing (Contract number of the key laboratory: 2012E10016).}
}
\begin{document}
\maketitle

\begin{abstract}

%In this work, we propose low complexity adaptive biased estimation algorithms, called group-based shrinkage estimators (GSEs), for parameter estimation and interference suppression scenarios with mechanisms to automatically adjust the shrinkage factors. The proposed estimation algorithms divide the target parameter vector into a number of groups and adaptively calculate one shrinkage factor for each group. GSE schemes improve the performance of the conventional least squares (LS) estimator in terms of mean-squared error (MSE), while requiring a very modest increase in complexity. An MSE analysis is presented which indicates the lower bounds of the GSE schemes with different {\color{red}group sizes}. We prove that our proposed schemes outperform the biased estimation with only one shrinkage factor and the best performance of GSE can be obtained with the maximum number of groups. Then, we consider an application of the proposed algorithms to single-carrier frequency domain equalization (SC-FDE) of direct-sequence ultra-wideband (DS-UWB) systems, in which the structured channel estimation (SCE) and frequency domain receiver are performed by using the GSE. The simulation results show that the proposed algorithms significantly outperform the conventional unbiased estimator in the analyzed scenarios.

In this work, we propose low-complexity adaptive biased estimation
algorithms, called group-based shrinkage estimators (GSEs), for parameter
estimation and interference suppression scenarios with mechanisms to
automatically adjust the shrinkage factors. The proposed estimation
algorithms divide the target parameter vector into a number of groups and
adaptively calculate one shrinkage factor for each group. GSE schemes
improve the performance of the conventional least squares (LS) estimator
in terms of the mean-squared error (MSE), while requiring a very modest
increase in complexity. An MSE analysis is presented which indicates the
lower bounds of the GSE schemes with different group sizes. We prove that
our proposed schemes outperform the biased estimation with only one
shrinkage factor and the best performance of GSE can be obtained with the
maximum number of groups. Then, we consider an application of the proposed
algorithms to single-carrier frequency-domain equalization (SC-FDE)
of direct-sequence ultra-wideband (DS-UWB) systems, in which the
structured channel estimation (SCE) algorithm and the frequency domain
receiver employ the GSE. The simulation results show that the proposed
algorithms significantly outperform the conventional unbiased estimator in
the analyzed scenarios.
\end{abstract}

\section{Introduction}

In this work, biased estimation algorithms are considered in two
common deterministic estimation scenarios in communications
engineering, which are parameter estimation and interference
suppression \cite{rodrigoadda2008}-\cite{MMS2010}.
%In the parameter estimation scenario such as the channel estimation task,
%the $M$-dimensional observation vector can be modeled as $\bm y =
%\bm X \bm h+ \bm n$, where $\bm X\in \mathbb C^{M\times L}$ is a
%given matrix and $\bm n\in \mathbb C^{M\times 1}$ represents the
%additive white Gaussian noise (AWGN) with zero mean and variance
%$\sigma_{n}^2$. The parameter vector $\bm h\in \mathbb C^{L\times
%1}$ is the target parameter vector to be estimated. In the
%interference suppression scenario such as the equalizer design, the
%$M$-dimensional observation vector can be modeled as $\bm z =\bm H
%\bm b +\bm n$, where $\bm H\in \mathbb C^{M\times N}$ is an unknown
%matrix that can represent the channel and/or the spreading codes and
%the $\bm n\in \mathbb C^{M\times 1}$ is the AWGN with zero mean and
%variance $\sigma_{n}^2$. In this scenario, a training sequence is
%transmitted and an equalizer whose parameters are organized in the
%vector $\bm w$ is estimated to recover the data vector $\bm b\in
%\mathbb C^{N\times 1}$.
It is known that under the assumption of AWGN, the least-square (LS)
algorithm can provide an efficient solution to these estimation
problems and will lead to minimum variance unbiased estimators
(MVUE). The unbiasness is usually considered as a good property for
an estimator because the expected value of unbiased estimators is
the true value of the unknown parameter \cite{smkay1993}. However,
in some scenarios the LS method is not directly related to the mean
square error (MSE) associated with the target parameter vector and
it has been found that a lower MSE can be achieved by adding an
appropriately chosen bias to the conventional LS estimators
\cite{Yceldar2005},\cite{KJ2011}. Note that some reduced-rank techniques also employ a bias to accelerate the convergence speed \cite{MWF}-\cite{jidf}.

A class of biased estimator that has been studied in recent years is
known as the biased estimators with a shrinkage factor
\cite{Yceldar2005}-\cite{myICASSP2011}. These biased estimation
algorithms have shown their ability to outperform the existing
unbiased estimators especially in low signal-to-noise ratios (SNR)
scenarios and/or with short data records \cite{Skay2008}. For these
biased estimators \cite{Yceldar2005}-\cite{myICASSP2011}, the
complexity is much lower than for MMSE algorithms because the
additional number of coefficients to be computed is only one. The
motivation for the group-based shrinkage estimator
(GSE) is to find a generalized estimator with a number of shrinkage
factors that can achieve a better performance and complexity
tradeoff than the biased estimator with only one shrinkage factor.

In the parameter estimation scenario,% the typical objective is to
%minimize the Euclidian norm of the estimation error $\|\bm
%h-\hat{\bm h}\|$, rather than minimizing the Euclidian norm of the
%reconstruction error $\|\bm y-\hat{\bm y}\|$ as in the LS algorithm,
%where $\hat{\bm y}=\bm X \hat{\bm h}$ can be considered as the
%transformed estimator \cite{Yceldar2005}.
some biased estimators have been proposed to achieve a smaller estimation error than the LS
solutions by removing the unbiasedness of the conventional
estimators with a shrinkage factor in the parameter estimation
scenario. The earliest shrinkage estimators that reduce the MSE over
MVUE include the well known James-Stein estimator \cite{wjames1961}
and the work of Thompson \cite{Thompson1968}. Some extensions of the
James-Stein estimator have been proposed in
\cite{MEBOCK1975}-\cite{JHMANTON1998}. In \cite{zbenhaim2007}, blind
minimax estimation (BME) techniques have been proposed, in which the
biased estimators were developed to minimize the worst case MSE
among all possible values of the target parameter vector within a parameter set. If a
spherical parameter set is assumed, the shrinkage estimator obtained
is named spherical BME (SBME) \cite{zbenhaim2007}.

For the interference suppression scenario, the biased estimators can
be employed to achieve a lower estimation error between the
estimated filter and the optimal linear LS estimator. %$\hat{\bm w}$ and the optimal linear LS estimator
%$\bm w_{\rm o}$.
%Although the typical target in this scenario is to
%minimize the squared estimation error $\|\bm b-\hat{\bm b}\|^2$
The major motivation for adopting the biased algorithms here is to
accelerate the convergence rate for the adaptive implementations and
provide a better performance with short training data support in
long filter scenarios \cite{myEW2011}.

To the best of our knowledge, biased estimators with shrinkage
factors are rarely implemented into real-world signal processing and
have not been considered in the frequency domain for communication
systems. One possible reason is that some assumptions required for
the signal model may not be satisfied. For example, in time-hopping
UWB (TH-UWB) systems, the multiple access interference (MAI) cannot
be accurately approximated by a Gaussian distribution for some
values of the the signal-to-interference-plus-noise ratio (SINR)
\cite{GDURISI2003}. Another possible reason is that the existing
shrinkage-based estimators usually require some statistical
information such as the noise variance and the norm of the actual
parameter vector. In our previous work \cite{myEW2011} and
\cite{myICASSP2011}, adaptive biased estimation algorithms with only
one shrinkage factor have been proposed to fulfill the tasks of
interference suppression and parameter estimation. In this work, a
novel biased estimation technique, named group-based shrinkage
estimators (GSE), is proposed. In this algorithm, the target
parameter vector is divided into several groups and one shrinkage
factor is calculated for each group. Least mean square (LMS)-based
adaptive estimation algorithms are then developed to calculate the
shrinkage factors. The GSE estimators are able to improve the
performance of the recursive least squares (RLS) algorithm that
recursively computes the LS estimator. In DS-UWB systems, the estimation tasks are usually very challenging because the environments include dense multipath. In this work, in order to test the proposed algorithms, we consider applications of
DS-UWB systems with SC-FDE. Specifically, we concentrate on the
channel estimation and interference suppression with the
proposed algorithms. The MSE performance of the proposed GSE schemes
is then analyzed, a lower bound of the MSE performance is derived
and the relationship between the number of groups and the lower
bound is set up. Simulations show that with an additional complexity
that is only linearly dependent on the size of the parameter vector
and the number of groups, the proposed biased GSE algorithms
outperform the conventional RLS algorithm in terms of MSE in low SNR
scenarios and/or with short data support. It should be noted that the proposed GSE estimator
can be employed for applications where a high estimation accuracy is
required. These include localization in wireless sensor networks
\cite{JINGTENG2012} and in dense cluttered environments with UWB
technology \cite{damien2008}. The proposed estimators can also be
employed into emergent multicast and broadcast systems \cite{MMS2010}, such as the orthogonal
frequency-division multiplexing (OFDM) based multi-user
multiple-input multiple-output (MIMO) systems as specified in the
IEEE 802.11ac standard and the 3GPP long-term-evolution (LTE)
systems.

The main contributions of this work are summarized as follows:

\begin{itemize}
\item {Novel GSE schemes are proposed to improve the performance of the frequency domain RLS algorithms in
the applications of parameter estimation and interference suppression in DS-UWB systems.}
\end{itemize}

\begin{itemize}
\item {LMS based adaptive algorithms are developed for both scenarios to adjust the shrinkage factors.}
\end{itemize}

\begin{itemize}
\item {The MSE analysis is carried out which indicates a lower bound of the proposed estimator and the relationship between the lower bound and the number of groups.}
\end{itemize}
\begin{itemize}
\item {The performance of the proposed biased estimators is examined in multiuser SC-FDE for DS-UWB systems with the IEEE 802.15.4a channel model, convolutional and low-density parity-check (LDPC) codes.}
\end{itemize}

The rest of this paper is structured as follows. In Section \ref{sec:sm}, we first review the LS solution for the parameter estimation scenario and present the structured channel estimation (SCE) problem in SC-FDE of DS-UWB systems. Then, the signal model of the frequency domain receiver design for DS-UWB systems that represents the interference suppression scenario, is presented. The proposed GSE scheme and its adaptive implementations for the parameter estimation scenario and the interference suppression scenario are developed in Section \ref{sec:PE} and Section \ref{sec:IS}, respectively. The MSE analysis is shown in Section \ref{sec:MSEanalysis}. The simulation results are shown in Section \ref{sec:simulations}. Section \ref{sec:conclusion} draws the conclusions.

%%%%%%%%%%%%%%%%%%%%%%%%%%%%%%%%%%%%%%%%%%%%%%%%%%%%%%%%%%%%%%%%%%%%%%%%%%%%%%%%%%%%%%%%%%%%%%%%%%%%%%%%%%%%%%%%%%%
\section{System model}
\label{sec:sm}
In this section, we introduce the channel estimation and receiver design tasks in the frequency-domain of DS-UWB systems with SC-FDE that represent the parameter estimation scenario and the interference suppression scenario, respectively.

\subsection{Problem statement for the parameter estimation scenario}
\label{sec:PE2}
%In the parameter estimation scenario such as the channel estimation task,
%the $M$-dimensional observation vector can be modeled as $\bm y =
%\bm X \bm h+ \bm n$, where $\bm X\in \mathbb C^{M\times L}$ is a
%given matrix and $\bm n\in \mathbb C^{M\times 1}$ represents the
%additive white Gaussian noise (AWGN) with zero mean and variance
%$\sigma_{n}^2$. The parameter vector $\bm h\in \mathbb C^{L\times
%1}$ is the target parameter vector to be estimated. In the
%interference suppression scenario such as the equalizer design, the
%$M$-dimensional observation vector can be modeled as $\bm z =\bm H
%\bm b +\bm n$, where $\bm H\in \mathbb C^{M\times N}$ is an unknown
%matrix that can represent the channel and/or the spreading codes and
%the $\bm n\in \mathbb C^{M\times 1}$ is the AWGN with zero mean and
%variance $\sigma_{n}^2$. In this scenario, a training sequence is
%transmitted and an equalizer whose parameters are organized in the
%vector $\bm w$ is estimated to recover the data vector $\bm b\in
%\mathbb C^{N\times 1}$.
The linear model for the parameter estimation scenario can be expressed as:
%%%%%%%%%%%%%%%%%%%%%%%%%%%%%%%%%%%%%%%%%%%%%%%%%%%%%%%%%%%%%%%%%%%%%%%%%%%%%%%%%%%%%%%%%%%%%%%%%%%%%%%%%%%%%%%%%%%
\begin{equation}
\bm y = \bm X \bm h+ \bm n,\label{equ:yxhn}
\end{equation} where the training data matrix $\bm X\in \mathbb C^{M\times L}$ and the received signal $\bm y \in \mathbb C^{M\times 1}$ are given, $\bm n$ is AWGN with zero mean and variance
$\sigma^2$. In this scenario, the typical target is to estimate the
parameter vector $\bm h\in \mathbb C^{L\times 1}$ that leads to the
minimum MSE. The MSE consists of the estimation variance and the
squared bias and is given by
%%%%%%%%%%%%%%%%%%%%%%%%%%%%%%%%%%%%%%%%%%%%%%%%%%%%%%%%%%%%%%%%%%%%%%%%%%%%%%%%%%%%%%%%%%%%%%%%%%%%%%%%%%%%%%%%%%%
\begin{equation*}
{\mathbb E}\{\|\bm h - \hat{\bm h}\|^2\}={\mathbb E}\{(\hat{\bm
h}-{\mathbb E}\{\hat{\bm h}\})^{H}(\hat{\bm h}-{\mathbb
E}\{\hat{\bm h}\})\}+[\|{\mathbb E}\{\hat{\bm h}\}-\bm
h\|^{2}], %\label{equ:expofMSE_biased}
\end{equation*}where ${\mathbb E}\{\cdot\}$ represents expectation of a random variable.

The conventional LS algorithm estimates the parameter by minimizing
%%%%%%%%%%%%%%%%%%%%%%%%%%%%%%%%%%%%%%%%%%%%%%%%%%%%%%%%%%%%%%%%%%%%%%%%%%%%%%%%%%%%%%%%%%%%%%%%%%%%%%%%%%%%%%%%%%%
\begin{equation}
{J}_{\rm LS}(\bm h)= \|\bm y-\bm X \bm h \|^2.
\end{equation} Assuming that the matrix $\bm X^{H}\bm X$ is a full rank matrix, the LS solution is given by
%%%%%%%%%%%%%%%%%%%%%%%%%%%%%%%%%%%%%%%%%%%%%%%%%%%%%%%%%%%%%%%%%%%%%%%%%%%%%%%%%%%%%%%%%%%%%%%%%%%%%%%%%%%%%%%%%%%
\begin{equation}
\hat {\bm h}_{\rm LS}= (\bm X^{H}\bm X)^{-1}\bm
X^{H}\bm y.
\end{equation}
Under the assumption of AWGN with zero mean and variance $\sigma^2$, the LS estimator is a MVUE that leads to a minimum MSE
%%%%%%%%%%%%%%%%%%%%%%%%%%%%%%%%%%%%%%%%%%%%%%%%%%%%%%%%%%%%%%%%%%%%%%%%%%%%%%%%%%%%%%%%%%%%%%%%%%%%%%%%%%%%%%%%%%%
\begin{equation*}
{\mathbb E}\{\|\bm h - \hat{\bm h}_{\rm LS}\|^2\}={\mathbb
E}\{(\hat{\bm h}_{\rm LS}-\bm h)^{H}(\hat{\bm h}_{\rm
LS}-\bm h)\}=var(\bm h,\hat{\bm h}_{\rm
LS}).%\label{equ:v_6}
\end{equation*} we define $v=var(\bm h,\hat{\bm h}_{\rm LS})=tr\{\sigma^2(\bm X^{H}\bm X)^{-1}\}$, where $tr\{\cdot\}$ represents the trace operator \cite{smkay1993}.

The objective of introducing the biased estimator in the parameter estimation scenario is to achieve a lower MSE than the unbiased estimator, which can be expressed as
%%%%%%%%%%%%%%%%%%%%%%%%%%%%%%%%%%%%%%%%%%%%%%%%%%%%%%%%%%%%%%%%%%%%%%%%%%%%%%%%%%%%%%%%%%%%%%%%%%%%%%%%%%%%%%%%%%%
\begin{equation}
{\mathbb E}\{\|\bm h - \hat{\bm h}_{\rm b}\|^2\}\leq{\mathbb
E}\{\|\bm h - \hat{\bm h}_{\rm LS}\|^2\}.
\end{equation}

Although the objective shown here is similar to MMSE algorithms that is to achieve an MSE as small as possible. It should be noted that the biased algorithm developed in this work adopts a different strategy from MMSE algorithms.

%In the following section, the channel estimation problem in SC-FDE of DS-UWB systems that has the same estimation model as in \eqref{equ:yxhn} will be introduced in detail.

\subsection{System model for the SCE: parameter estimation scenario}
\label{sec:systemmodelsce}
%%%%%%%%%%%%%%%%%%%%%%%%%%%%%%%%%%%%%%%%%%%%%%%%%%%%%%%%%%%%%%%%%%%%%%%%%%%%%%%%%%%%%%%%%%%%%%%%%%%%%%%%
\begin{figure*}[htb]
\begin{minipage}[b]{1.0\linewidth}
  \centering
  \centerline{\epsfig{figure=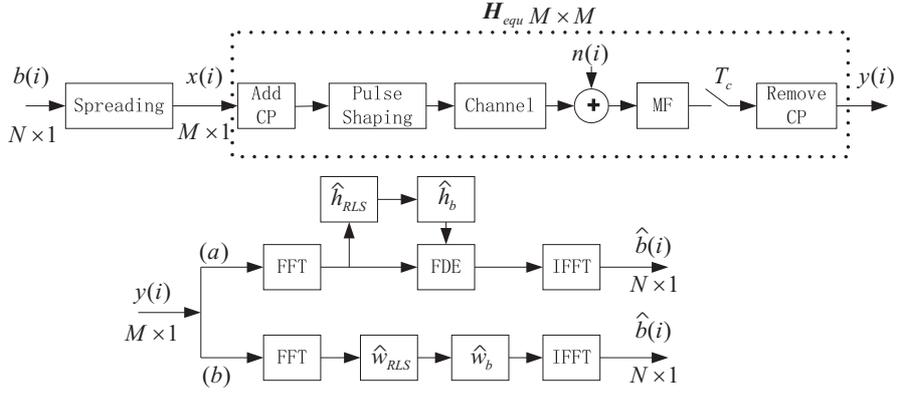,scale=0.75}}
\end{minipage}
\caption{Block diagram of SC-FDE schemes in DS-UWB systems: (a) Parameter estimation scenario,(b) Interference suppression scenario.}
\label{fig:blockdiag_sys}
\end{figure*}
Here, we consider the channel estimation problem of a synchronous downlink block-by-block transmission DS-UWB system based on SC-FDE with $K$ users. The block diagram of the parameter estimation scenario is shown as branch (a) in Fig. \ref{fig:blockdiag_sys}. For notational simplicity, we assume that a $N_{c}$-by-$1$ Walsh spreading code $\bm s_{k}$ is assigned to the $k$-th user. The spreading gain is $N_{c}=T_{s}/T_{c}$, where $T_{s}$ and $T_{c}$ denote the symbol duration and chip duration, respectively. At each time instant $i$, a data vector $\bm b_{k}(i)\in \mathbb C^{N\times 1}$ is transmitted by the $k$-th user. We define the signal after spreading as $\bm x_{k}(i)=\bm D_{k}\bm
b_{k}(i)$, where the block diagonal matrix $\bm D_{k}\in \mathbb C^{M\times N}$ ($M=N \cdot N_{c}$) performs the spreading of the data block and its first column is constructed by the spreading code  $\bm s_{k}$ zero-padded to the length of $M$. In order to prevent inter block interference (IBI), a cyclic-prefix (CP) is added and the length of the CP is assumed to be larger than the length of the channel impulse response (CIR). With the insertion of the CP at the transmitter and its removal at the receiver, the equivalent channel is denoted as a circulant Toeplitz matrix $\bm H_{\rm equ}\in \mathbb C^{M\times M}$, whose first column is composed of a vector $\bm h_{\rm equ}$ zero-padded to length $M$, where $\bm h_{\rm equ}=[h(0), h(1),\dots, h(L-1)]$ is the equivalent CIR. At the receiver, a chip matched-filter (CMF) is applied and the received sequence is then sampled at chip-rate and organized in an $M$-dimensional vector. This signal then goes through the discrete Fourier transform (DFT). The frequency-domain received signal is given by
%%%%%%%%%%%%%%%%%%%%%%%%%%%%%%%%%%%%%%%%%%%%%%%%%%%%%%%%%%%%%%%%%%%%%%%%%%%%%%%%%%%%%%%%%%%%%%%%%%%%%%%%
\begin{equation}
\bm y(i)=\bm F \bm H_{\rm equ}\sum_{k=1}^{K}\bm
x_{k}(i)+\bm F\bm n(i)=\bm{\Lambda}_{\rm H}\bm
F\sum_{k=1}^{K}\bm x_{k}(i)+\bm F\bm n(i),
\label{eq:scezi}
\end{equation} where $\bm n(i)$ represents the AWGN, $\bm{\Lambda}_{\rm H}=\bm F\bm H_{\rm equ}\bm F^{H}$ is a diagonal matrix whose diagonal vector is defined as $\tilde{\bm h}$ and its $a$-th entry is given by $\tilde{h}_{a}=\sum_{l=0}^{L-1}h_{l}\cdot {\rm exp}\{-j(2\pi/M)a\hspace{0.15em}l\}$, where $a$ $\in$ $\{0,1,2,\dots,M-1\}$. $\bm F\in \mathbb C^{M\times M}$ represents the DFT matrix and its $(a,b)$-th entry is ${\bm F}_{a,b}$$=(1/\sqrt{M}){\rm exp}\{-j(2\pi/M)a\hspace{0.15em}b\},$ where $a,b$ $\in$ $\{0,1,2,\dots,M-1\}$. By defining a matrix $\bm F_{M,L}\in \mathbb C^{M\times L}$ that contains the first $L$ columns of the DFT matrix $\bm F$, we obtain the following relationship

%%%%%%%%%%%%%%%%%%%%%%%%%%%%%%%%%%%%%%%%%%%%%%%%%%%%%%%%%%%%%%%%%%%%%%%%%%%%%%%%%%%%%%%%%%%%%%%%%%%%%%%%
\begin{equation}
\tilde{\bm h}=\sqrt{M}\bm F_{M,L}\bm h_{\rm equ},
\label{eq:tildeh}
\end{equation}
In unstructured channel estimation (UCE), the vector $\tilde{\bm h}\in \mathbb C^{M\times 1}$ is directly estimated, while in the structured channel estimation (SCE), the fact that $L<M$ is taken into account and the vector $\bm h_{\rm equ}\in \mathbb C^{L\times 1}$ is the parameter vector to be estimated. The concept of SCE was proposed in \cite{mmorelli2005}, where the SCE shows a better performance than the UCE. In \cite{myiet}, adaptive MMSE detection schemes for SC-FDE in multi-user DS-UWB systems based on SCE are developed, where the estimated $\bm h_{\rm equ}$ is adaptively calculated based on RLS, least-mean squares (LMS) and the conjugate gradient (CG) algorithm for the detection and the RLS version performs the best. The purpose of developing biased estimation in this scenario is to further improve the performance of the RLS algorithm in terms of the MSE.

We consider user 1 as the desired user and omit the subscript of this user for simplicity. Note that the frequency domain received signal can be expressed as
%%%%%%%%%%%%%%%%%%%%%%%%%%%%%%%%%%%%%%%%%%%%%%%%%%%%%%%%%%%%%%%%%%%%%%%%%%%%%%%%%%%%%%%%%%%%%%%%%%%%%%%%
\begin{equation}
\bm y(i)=\sqrt{M}\bm \Delta(i)\bm F_{M,L}\bm h_{\rm
equ}+\bm n_{\rm e}(i), \label{eq:scezifinal}
\end{equation} where we define a diagonal matrix $\bm \Delta(i)={\rm diag}[\bm F\bm x(i)]$, the noise and interference vector $\bm n_{\rm e}(i)$ consists of the MAI and the noise and is assumed to be AWGN. As shown in \eqref{eq:scezifinal}, the SCE problem is an implementation example of the parameter estimation problem where a given matrix  $\bm X(i)\in \mathbb C^{M\times L}$ is defined as $\bm X(i)=\sqrt{M}\bm \Delta(i)\bm F_{M,L}$. The LS
solution of $\bm h_{\rm equ}$ is given by
%%%%%%%%%%%%%%%%%%%%%%%%%%%%%%%%%%%%%%%%%%%%%%%%%%%%%%%%%%%%%%%%%%%%%%%%%%%%%%%%%%%%%%%%%%%%%%%%%%%%%%%%
\begin{equation}
{\bm h}_{\rm equ,LS}(i)=\bm R_{\rm h}^{-1}(i)\bm p_{\rm
h}(i)
\end{equation} where
$\bm R_{\rm h}(i)= \sum_{j=1}^{i}\lambda^{i-j}\bm
F^{H}_{M,L}\bm \Delta^{H}(j)\bm \Delta(j)\bm F_{M,L}$,
$\bm p_{\rm h}(i)=\sum_{j=1}^{i}\lambda^{i-j}\bm
F^{H}_{M,L}\bm \Delta^{H}(j)\bm z(j)$ and $\lambda$ is the
forgetting factor. Then the LS solution can be computed recursively
by the following RLS algorithm \cite{mmorelli2005}
%%%%%%%%%%%%%%%%%%%%%%%%%%%%%%%%%%%%%%%%%%%%%%%%%%%%%%%%%%%%%%%%%%%%%%%%%%%%%%%%%%%%%%%%%%%%%%%%%%%%%%%%
\begin{equation}
\hat {\bm h}_{\rm RLS}(i+1)=\hat {\bm h}_{\rm RLS}(i)+
\bm R_{\rm h}^{-1}(i)\bm F^{H}_{M,L}\bm
\Delta^{H}(i)\bm e_{\rm h}(i),\label{equ:rlsforsce}
\end{equation} where $\bm e_{\rm h}(i)=\bm z(i)-\bm \Delta(i)\bm F_{M,L}\hat
{\bm h}_{\rm RLS}(i)$ is the $M$-dimensional error vector.

In Section \ref{sec:PE}, a novel biased estimation algorithm called group-based shrinkage estimator (GSE) is incorporated into the unbiased LS estimator that is able to improve the estimation performance in terms of the MSE.

\subsection{System model for the frequency domain receiver design: interference suppression scenario}
\label{sec:systemmodelis}
%In this section, the proposed GSE schemes will be employed in the interference suppression scenario to achieve a lower estimation error between the estimated receiver $\hat{\bm w}$ and the optimum tap-weight vector of the receiver $\bm w_{\rm o}$ (optimum in the MSE sense).
The block diagram of the interference suppression scenario is shown as branch (b) in Fig. \ref{fig:blockdiag_sys}. For each time instant $i$, an $N$-dimensional data vector $\bm b_{k}(i)=[b_{k}^{(1)}(i),\dots,b_{k}^{(N)}(i)]^T$ is transmitted by the $k$-th user. After the spreading, the $M$-dimensional transmit signal is given by
%%%%%%%%%%%%%%%%%%%%%%%%%%%%%%%%%%%%%%%%%%%%%%%%%%%%%%%%%%%%%%%%%%%%%%%%%%%%%%%%%%%%%%%%%%%%%%%%%%%%%%%%
\begin{equation}
\bm x_{k}(i)=\bm S_{k}\bm b_{k,{\rm e}}(i),
\end{equation} where the spreading matrix $\bm S_{k}\in \mathbb C^{M\times M}$, $M=N \times N_{c}$, is a circulant Toeplitz matrix and its first column consists of the spreading codes and zero-padding \cite{MingXianchang2006}.
%\begin{equation*}
%{\renewcommand{\baselinestretch}{0.3} \footnotesize{
% \bm
%S_{k}=\begin{bmatrix}
% s_{k}(1)     &             &               & s_{k}(2)      \\
% s_{k}(2)     & s_{k}(1)    &               & \vdots        \\
% \vdots       & s_{k}(2)    &               & s_{k}(N_{c})  \\
%s_{k}(N_{c})  & \vdots      &\ddots         &              \\
%             & s_{k}(N_{c})&\ddots \\
%             &             &\ddots \\
%              &             &               & s_{k}(1)      \\
%\end{bmatrix},}}
%\end{equation*}
%%%%%%%%%%%%%%%%%%%%%%%%%%%%%%%%%%%%%%%%%%%%%%%%%%%%%%%%%%%%%%%%%%%%%%%%%%%%%%%%%%%%%%%%%%%%%%%%%%%%%%%%
The equivalent $M$-dimensional expanded data vector is
%%%%%%%%%%%%%%%%%%%%%%%%%%%%%%%%%%%%%%%%%%%%%%%%%%%%%%%%%%%%%%%%%%%%%%%%%%%%%%%%%%%%%%%%%%%%%%%%%%%%%%%%
\begin{equation*}
\bm b_{k,{\rm e}}(i)=[b_{k}^{(1)}(i),\bm 0_{N_{c}-1},b_{k}^{(2)}(i),\bm
0_{N_{c}-1},\cdots,b_{k}^{(N)}(i),\bm 0_{N_{c}-1}]^{T},
\end{equation*} where $(\cdot)^T$ is the transpose. Using this signal expression we can obtain a simplified frequency domain receiver design. At the receiver, a CMF is applied and the received sequence is then sampled at chip-rate and organized in an $M$-dimensional vector. After the DFT, the received signal is given by
%%%%%%%%%%%%%%%%%%%%%%%%%%%%%%%%%%%%%%%%%%%%%%%%%%%%%%%%%%%%%%%%%%%%%%%%%%%%%%%%%%%%%%%%%%%%%%%%%%%%%%%%
\begin{equation}
\bm z(i)=\bm F \bm y(i)=\sum_{k=1}^{K}\bm F \bm H_{\rm
equ}\bm S_{k}\bm b_{k,{\rm e}}(i)+\bm F\bm n(i),
\end{equation} where $\bm n(i)$ is the AWGN and $\bm F\in \mathbb C^{M\times M}$ represents the DFT matrix. Since both $\bm H_{\rm equ}$ and $\bm S_{k}$ are circulant Toeplitz matrices, their product also has the circulant Toeplitz form. This feature makes $\bm{\Lambda}_{k}=\bm F \bm H_{\rm equ}\bm S_{k} \bm F^{H}$ a diagonal matrix. Hence, we have
%%%%%%%%%%%%%%%%%%%%%%%%%%%%%%%%%%%%%%%%%%%%%%%%%%%%%%%%%%%%%%%%%%%%%%%%%%%%%%%%%%%%%%%%%%%%%%%%%%%%%%%%
\begin{equation}
%\begin{split}
\bm z(i)%&=\sum_{k=1}^{K}\mathbf F \mathbf H_{\rm equ}\mathbf S_{k} \mathbf
%F^{H}\mathbf F\mathbf b_{k,{\rm e}}(i)+\mathbf F\mathbf
%n(i)\\
=\sum_{k=1}^{K}\bm{\Lambda}_{k} \bm F\bm b_{k,{\rm e}}(i)+\bm
F\bm n(i).
%\end{split}
\end{equation}
%%%%%%%%%%%%%%%%%%%%%%%%%%%%%%%%%%%%%%%%%%%%%%%%%%%%%%%%%%%%%%%%%%%%%%%%%%%%%%%%%%%%%%%%%%%%%%%%%%%%%%%%
We can further expand $\bm F\bm b_{k,{\rm e}}(i)$ as \cite{MingXianchang2006}
%%%%%%%%%%%%%%%%%%%%%%%%%%%%%%%%%%%%%%%%%%%%%%%%%%%%%%%%%%%%%%%%%%%%%%%%%%%%%%%%%%%%%%%%%%%%%%%%%%%%%%%%
\begin{equation}
\bm F\bm b_{k,{\rm e}}(i)=(1/\sqrt {N_{c}})\bm I_{\rm e}\bm F_{N}\bm b_{k}(i),
\end{equation}
where $\bm F_{N}\in \mathbb C^{N\times N}$ denotes the DFT matrix and $\bm I_{\rm e}\in \mathbb C^{M\times N}$ is structured as
%%%%%%%%%%%%%%%%%%%%%%%%%%%%%%%%%%%%%%%%%%%%%%%%%%%%%%%%%%%%%%%%%%%%%%%%%%%%%%%%%%%%%%%%%%%%%%%%%%%%%%%%
\begin{equation}
\bm I_{\rm e}=[\underbrace{\bm I_{N},\cdots,\bm I_{N}}_{N_{c}}]^{T}. \end{equation} where $\bm I_{N}$ denotes the $N$-by-$N$ identity matrix. Finally, the frequency domain received signal
$\bm z(i)$ is given by
%%%%%%%%%%%%%%%%%%%%%%%%%%%%%%%%%%%%%%%%%%%%%%%%%%%%%%%%%%%%%%%%%%%%%%%%%%%%%%%%%%%%%%%%%%%%%%%%%%%%%%%%
\begin{equation}
\bm z(i)=\sum_{k=1}^{K}(1/\sqrt {N_{c}})\bm{\Lambda}_{k}\bm I_{\rm
e}\bm F_{N}\bm b_{k}(i)+\bm F\bm n(i). \label{eq:fdinputzi}
\end{equation} Note that the expression in \eqref{eq:fdinputzi} is an implementation example of the interference suppression scenario where the unknown matrix for each time instant is given by $\bm H(i)=(1/\sqrt {N_{c}})\bm{\Lambda}_{k}\bm I_{\rm e}\bm F_{N}$.
%%%%%%%%%%%%%%%%%%%%%%%%%%%%%%%%%%%%%%%%%%%%%%%%%%%%%%%%%%%%%%%%%%%%%%%%%%%%%%%%%%%%%%%%%%%%%%%%%%%%%%%%
To fulfill the interference suppression task, an MMSE filter $\bm {W}(i)\in \mathbb C^{M\times N}$ can be developed
via the following cost function:
%%%%%%%%%%%%%%%%%%%%%%%%%%%%%%%%%%%%%%%%%%%%%%%%%%%%%%%%%%%%%%%%%%%%%%%%%%%%%%%%%%%%%%%%%%%%%%%%%%%%%%%%
\begin{equation}
{J}_{\rm MSE}(i)={\mathbb E}\big\{\left\|\bm b(i)-\bm {F}_{N}^{H}\bm
W^{H}(i)\bm z(i)\right\|^{2}\big\}. \label{eq:mmse}
\end{equation}
%%%%%%%%%%%%%%%%%%%%%%%%%%%%%%%%%%%%%%%%%%%%%%%%%%%%%%%%%%%%%%%%%%%%%%%%%%%%%%%%%%%%%%%%%%%%%%%%%%%%%%%%
The MMSE solution is given by \cite{myiet}
\begin{equation}
\bm W_{\rm
MMSE}%&=\left(\frac{1}{N_{c}}\sum_{k=1}^{K}\mathbf{\Lambda}_{k}\mathbf I_{\rm
%e}\mathbf I_{\rm e}^{H}\mathbf{\Lambda}_{k}^{H}+\sigma^{2}\mathbf
%I\right)^{-1}\frac{\mathbf{\Lambda}_{k}\mathbf I_{\rm e}}{\sqrt
%{N_{c}}}
=\bm V\bm I_{\rm e},
\end{equation} where the matrix $\bm V\in \mathbb C^{M\times M}$ is
\begin{equation}
\bm V=\frac{1}{\sqrt
{N_{c}}}\left(\frac{1}{N_{c}}\sum_{k=1}^{K}\bm{\Lambda}_{k}\bm I_{\rm
e}\bm I_{\rm e}^{H}\bm{\Lambda}_{k}^{H}+\sigma^{2}\bm
I_{M}\right)^{-1}\bm{\Lambda}_{k},
\end{equation}where $\bm I_{M}\in \mathbb C^{M\times M}$ denotes the identity matrix. Note that the matrix $\bm V$ consists of $N_{c}$ times $N_{c}$ diagonal matrices $\bm V_{ij}\in \mathbb C^{N\times N}$, where $i,j$ $\in$ $\{1,N_{c}\}$. Hence, we take a closer look at the product of $\bm V$ and $\bm I_{\rm e}$:
\begin{equation*}
\begin{split}
&
 \bm V\bm I_{e}=\begin{bmatrix}
 \bm V_{1,1}         & \bm V_{1,2}        & \dots       &\bm V_{1,N_{c}}      \\
 \bm V_{2,1}         & \bm V_{2,2}        & \dots       &\bm V_{2,N_{c}}        \\
 \vdots                  &  \vdots                & \vdots      & \vdots  \\
 \bm V_{N_{c},1}     & \bm V_{N_{c},2}    & \dots       & \bm V_{N_{c},N_{c}}      \\
\end{bmatrix}\begin{bmatrix}
 \bm I_{N}  \\
 \bm I_{N}  \\
 \vdots         \\
 \bm I_{N}  \\
\end{bmatrix}\\
&=\begin{bmatrix}
 \sum_{j=1}^{N_{c}}\bm V_{1,j}  \\
 \sum_{j=1}^{N_{c}}\bm V_{2,j}  \\
 \vdots         \\
 \sum_{j=1}^{N_{c}}\bm V_{N_{c},j}  \\
\end{bmatrix}
=\begin{bmatrix}
\hat {\bm W}_{1}        &              &             &                  \\
                    & \hat {\bm W}_{2} &             &                  \\
                    &              &\ddots       &                  \\
                    &              &             & \hat {\bm W}_{N_{c}}     \\
\end{bmatrix}\begin{bmatrix}
 \bm I_{N}  \\
 \bm I_{N}  \\
 \vdots         \\
 \bm I_{N}  \\
\end{bmatrix}=\hat {\bm W}\bm I_{\rm e},
\end{split}
\end{equation*} where $\hat {\bm W}_{i}=\sum_{j=1}^{N_{c}}\bm V_{i,j}$, $i=1,\dots,N_{c}$, are diagonal matrices. Hence, the product of $\bm V$ and $\bm I_{\rm e}$ can be converted into a product
of a diagonal matrix $\hat {\bm W}\in \mathbb C^{M\times M}$ ($M=N \times N_{c}$) and $\bm I_{\rm e}$, where the diagonal entries of $\hat {\bm W}$ are $\hat {w}_{l}$, $l=1,\dots,M$, and equal the sum of all entries in the $l$-th row of matrix $\bm V$. Finally, we express the MMSE design as
\begin{equation}
\bm W_{\rm MMSE}=\hat{\bm W}\bm I_{\rm e}={\rm diag}(\hat {\bm
w}_{\rm e}) \bm I_{\rm e}, \label{eq:vectormmse}
\end{equation} where $\hat {\bm w}_{\rm e}=(\hat {w}_{1},\hat {w}_{2},\dots,\hat {w}_{M})$ is an equivalent filter with $M$ taps.

The expression shown in \eqref{eq:vectormmse} enables us to design an $M$-dimensional receive filter rather than an $M$-by-$N$ matrix form receive filter. The estimated data vector can be expressed as
%%%%%%%%%%%%%%%%%%%%%%%%%%%%%%%%%%%%%%%%%%%%%%%%%%%%%%%%%%%%%%%%%%%%%%%%%%%%%%%%%%%%%%%%%%%%%%%%%%%%%%%%
\begin{equation}
\hat{\bm b}(i)=\bm {F}_{N}^{H}\bm I_{\rm e}^{H}\hat{\bm
W}^{H}(i)\bm z(i)=\bm {F}_{N}^{H}\bm I_{\rm e}^{H}\hat{\bm
Z}(i)\hat{\bm w}(i),
\end{equation} where $\hat{\bm
Z}(i)={\rm diag}(\bm z(i))$ and $\hat{\bm w}(i)=\hat{\bm w}_{\rm
e}^{*}(i)$ is the weight vector of the adaptive receiver. Since $\bm
{F}_{N}$ and $\bm I_{\rm e}$ are fixed, we consider the equivalent
$N$-by-$M$ received data matrix as $\bm Y(i)=\bm {F}_{N}^{H}\bm
I_{\rm e}^{H}\hat{\bm Z}(i),$
%%%%%%%%%%%%%%%%%%%%%%%%%%%%%%%%%%%%%%%%%%%%%%%%%%%%%%%%%%%%%%%%%%%%%%%%%%%%%%%%%%%%%%%%%%%%%%%%%%%%%%%%%
%\begin{equation}
%\mathbf Y(i)=\mathbf {F}_{N}^{H}\mathbf I_{\rm e}^{H}\hat{\mathbf
%Z}(i),\label{eq:nbymreceivedmatrix}
%\end{equation}
%%%%%%%%%%%%%%%%%%%%%%%%%%%%%%%%%%%%%%%%%%%%%%%%%%%%%%%%%%%%%%%%%%%%%%%%%%%%%%%%%%%%%%%%%%%%%%%%%%%%%%%%%
and express the estimated data vector as $\hat{\bm b}(i)=\bm Y(i)\hat{\bm w}(i)$.

%%%%%%%%%%%%%%%%%%%%%%%%%%%%%%%%%%%%%%%%%%%%%%%%%%%%%%%%%%%%%%%%%%%%%%%%%%%%%%%%%%%%%%%%%%%%%%%%%%%%%%%%%
\subsection{LS solution and adaptive RLS algorithm for the interference suppression scenario}
Here, we detail the LS and RLS designs for the frequency domain multiuser receiver $\hat{\bm w}$. The cost function for the development of the LS estimation is given by
%%%%%%%%%%%%%%%%%%%%%%%%%%%%%%%%%%%%%%%%%%%%%%%%%%%%%%%%%%%%%%%%%%%%%%%%%%%%%%%%%%%%%%%%%%%%%%%%%%%%%%%%
\begin{equation}
J_{\rm LS}= \|\bm b - \bm Y \bm w\|^{2}.
\end{equation} The LS design of the linear receiver can be expressed as
%%%%%%%%%%%%%%%%%%%%%%%%%%%%%%%%%%%%%%%%%%%%%%%%%%%%%%%%%%%%%%%%%%%%%%%%%%%%%%%%%%%%%%%%%%%%%%%%%%%%%%%%
\begin{equation}
\hat{\bm w}_{\rm LS}= (\bm Y^{H}\bm Y)^{-1}\bm Y^{H} \bm b=
\bm R_{\rm LS}^{-1}\bm p_{\rm LS},\label{equ:wls_6}
\end{equation} where the matrix $\bm R_{\rm LS}$ is defined as $\bm Y^{H}\bm Y$ and $\bm p_{\rm LS}$ represents the vector $\bm Y^{H} \bm b$.
Note that, the data vector can be expressed as
%%%%%%%%%%%%%%%%%%%%%%%%%%%%%%%%%%%%%%%%%%%%%%%%%%%%%%%%%%%%%%%%%%%%%%%%%%%%%%%%%%%%%%%%%%%%%%%%%%%%%%%%
\begin{equation}
\bm b = \bm Y \bm w_{\rm o}+\bm {\epsilon}_{\rm o},\label{eq:bywe}
\end{equation} where $\bm {\epsilon}_{\rm o}$ is the measurement
error vector and $\bm w_{\rm o}$ is the optimum tap-weight vector of the receiver (optimum in the MSE sense). Assuming that $\bm {\epsilon}_{\rm o}$ is white and Gaussian with zero
mean and covariance of $\sigma_{e}^2\bm I_{N}$, then the LS solution in
\eqref{equ:wls_6} is a MVUE \cite{haykin}.
%%%%%%%%%%%%%%%%%%%%%%%%%%%%%%%%%%%%%%%%%%%%%%%%%%%%%%%%%%%%%%%%%%%%%%%%%%%%%%%%%%%%%%%%%%%%%%%%%%%%%%%%%
%\begin{equation}
%E[\hat{\mathbf w}_{\rm LS}]= \mathbf w_{\rm o},
%\end{equation}which indicates that the LS estimator of the equalizer is an unbiased
%estimator. Assuming that $\epsilon_{\rm o}$ is white and Gaussian with zero
%mean and covariance of $\sigma_{e}^2\mathbf I$.
Now, let us have a look at the following MSE:
%%%%%%%%%%%%%%%%%%%%%%%%%%%%%%%%%%%%%%%%%%%%%%%%%%%%%%%%%%%%%%%%%%%%%%%%%%%%%%%%%%%%%%%%%%%%%%%%%%%%%%%%%%%%%%%%%%
\begin{equation}
{\mathbb E}\{\|\bm w_{\rm o} - \hat{\bm w}_{\rm LS}\|^2\}={\mathbb E}\{(\bm
w_{\rm o} - \hat{\bm w}_{\rm LS})^{H}(\bm w_{\rm o} - \hat{\bm
w}_{\rm LS})\}=var(\bm w_{\rm o},\hat{\bm w}_{\rm LS}).
\end{equation} Defining $v_{w}=var(\bm w_{\rm o},\hat{\bm w}_{\rm LS})$, we
have \cite{smkay1993}
\begin{equation}
v_{w}=tr\{\sigma_{e}^2(\bm Y^{H}\bm Y)^{-1}\}, \label{equ:v2_6}
\end{equation}where $\sigma_{e}^2$ is the variance of the measurement
error.

In the interference suppression scenario, it is possible to introduce the
biased estimation to reduce the MSE between the optimal receive filter $\bm
w_{\rm o}$ and the LS estimator $\hat{\bm w}_{\rm LS}$. Note that, for the
interference suppression scenario, the typical objective is to minimize the
overall performance criterion which is determined as $\mathbb {E}\{\|\bm b-\hat
{\bm b}\|^2\}$, rather than to minimize ${\mathbb E}\{\|\bm w_{\rm o} -
\hat{\bm w}\|^2\}$. The main motivation to introduce the bias in the
interference suppression scenario is to provide an initial improvement for the
overall performance when the adaptive filtering techniques are employed and the
training data are limited. This can also help with tracking problems and with
robustness against interference.

The LS solution of the receiver can be computed recursively by the RLS adaptive
algorithm. We employ the RLS update equation that is proposed in \cite{myiet}
%%%%%%%%%%%%%%%%%%%%%%%%%%%%%%%%%%%%%%%%%%%%%%%%%%%%%%%%%%%%%%%%%%%%%%%%%%%%%%%%%%%%%%%%%%%%%%%%%%%%%%%%
\begin{equation}
\hat{\bm w}_{\rm RLS}(i+1)=\hat{\bm w}_{\rm RLS}(i)+\bm R_{\rm w}^{-1}(i)\bm
Y^{H}(i)\bm e_{\rm aw}(i),\label{eq:darlsw}
\end{equation}  where $\bm R_{\rm w}(i)=
\sum_{j=1}^{i}\lambda_{w}^{i-j}\bm Y^{H}(j)\bm Y(j)$ and $\bm
e_{\rm aw}(i)=\bm b(i)-\bm Y(i)\hat{\bm w}(i)$. Note that $\bm
R_{\rm w}(i)$ is an $M$-by-$M$ symmetric sparse matrix in which the number of
nonzero elements equals $MN_{c}$. Hence, the complexity of each adaptation by
using this algorithm is $\mathcal{O}(MN_{c}^{2})$.

%%%%%%%%%%%%%%%%%%%%%%%%%%%%%%%%%%%%%%%%%%%%%%%%%%%%%%%%%%%%%%%%%%%%%%%%%%%%%%%%%%%%%%%%%%%%%%%%%%%%%%%%%%%%%%%%%%%
\section{Proposed GSE for parameter estimation scenario}
\label{sec:PE}

%In order to develop the adaptive biased algorithms in parameter estimation scenarios, the SCE task in the frequency domain of DS-UWB systems will be considered. Firstly, the problem statement of the biased estimation for parameter estimation scenario will be explained with a brief review of the conventional LS algorithm and the expression of the MSE. Then, the system model for the SCE implementation in the frequency domain of DS-UWB systems is described which is followed by the proposed biased estimation algorithms.

\subsection{Proposed GSE: Optimal Solution}
\label{sec:Optgse_sce}
It is known that the biased estimator with a shrinkage factor can be expressed as
%%%%%%%%%%%%%%%%%%%%%%%%%%%%%%%%%%%%%%%%%%%%%%%%%%%%%%%%%%%%%%%%%%%%%%%%%%%%%%%%%%%%%%%%%%%%%%%%%%%%%%%%%%%%%%%%%%%
\begin{equation}
\hat{\bm h}_{\rm b} = (1+\alpha)\hat{\bm h}_{\rm
LS},\label{equ:hbexpressionwithsf}
\end{equation} where $\hat{\bm h}_{\rm LS}\in \mathbb C^{L\times 1}$ is the LS estimator of the parameter vector and $\hat{\bm h}_{\rm b}\in \mathbb C^{L\times 1}$ is the biased estimator with a shrinkage factor, $\alpha$ is a real-valued variable and $(1+\alpha)$ is defined as the real-valued shrinkage factor that is larger than 0 but smaller than 1 (i.e., $-1<\alpha<0$).

Actually, for the parameter estimation scenario, the MMSE estimators with the following expression can also be considered as a biased estimator,
%%%%%%%%%%%%%%%%%%%%%%%%%%%%%%%%%%%%%%%%%%%%%%%%%%%%%%%%%%%%%%%%%%%%%%%%%%%%%%%%%%%%%%%%%%%%%%%%%%%%%%%%%%%%%%%%%%%
\begin{equation}
\hat{\bm h}_{\rm MMSE} = \bm A\hat{\bm h}_{\rm
LS},\label{equ:HMMSE}
\end{equation}where $\hat{\bm h}_{\rm MMSE}\in \mathbb C^{L\times 1}$ and $\bm A \in \mathbb C^{L\times L}$ is a full-rank matrix. As in \cite{MBiguesh2006} and \cite{JJbeek1995}, such MMSE channel estimators are developed for MIMO and OFDM systems, respectively. Although these MMSE estimators can achieve a much lower MSE than the LS estimator especially in low SNR regime, they experience much higher complexity than the biased estimator with only one shrinkage factor. In \cite{MBiguesh2006}, the proposed scaled LS channel estimator can be considered as a biased estimator with only one shrinkage factor, which outperforms the conventional LS estimator while it requires a much lower complexity than the MMSE estimator. The basic idea of the following proposed group-based shrinkage estimator (GSE) is to find a solution with a better tradeoff between the complexity and the performance than the MMSE estimator and the biased estimator with only one shrinkage factor.

The proposed GSE can be expressed as follows:
%%%%%%%%%%%%%%%%%%%%%%%%%%%%%%%%%%%%%%%%%%%%%%%%%%%%%%%%%%%%%%%%%%%%%%%%%%%%%%%%%%%%%%%%%%%%%%%%%%%%%%%%%%%%%%%%%%%
\begin{equation}
\hat{\bm h}_{\rm b} =\begin{bmatrix}
 (1+\alpha_{1})\hat{h}_{\rm LS}(1)    \\
 \vdots      \\
(1+\alpha_{1})\hat{h}_{\rm LS}(\frac{L}{S})\\
 (1+\alpha_{2})\hat{h}_{\rm LS}(\frac{L}{S}+1)              \\
  \vdots      \\
  (1+\alpha_{2})\hat{h}_{\rm LS}(\frac{2L}{S})\\
   \vdots      \\
    (1+\alpha_{S})\hat{h}_{\rm LS}(\frac{(S-1)L}{S}+1)              \\
  \vdots      \\
  (1+\alpha_{S})\hat{h}_{\rm LS}(L)\\
\end{bmatrix}
= \hat{\bm H}_{\rm LS} ({\bf 1}_{S}+\boldsymbol \alpha)%\left(\begin{bmatrix}
\label{equ:proposedhb}
\end{equation}where $\hat{\bm H}_{\rm LS}\in \mathbb C^{L\times S}$ is a block diagonal matrix that is constructed from
the elements of $\hat{\bm h}_{\rm LS}$ as well as zeros, $S$ is the number of groups, we define the $S$-dimensional column vectors ${\bf 1}_{S} =[1, 1,\dots, 1]^{T} $ and $\boldsymbol \alpha =[\alpha_{1} , \alpha_{2},\dots, \alpha_{S} ]^{T} $.
The scalar $\alpha_s$ is a real-valued variable and $(1+\alpha_{s})$ is defined as the shrinkage
factor for the $s$-th group of coefficients that is larger than 0 but smaller than 1, where $s=1,2,\dots,S$.
Here we propose to use a uniform group size for the $L$-dimensional parameter vector, hence the size of each group is $L/S$.
If the length of the parameter vector divided by the group size is not an integer, we can perform zero-padding in the parameter estimation vector to fulfill this requirement. If any statistical knowledge of the parameter vector is given, the group size could be different from each other. But this approach will introduce a higher complexity because we need to select the size of each group and choose a suitable one. In this work, for notational simplicity, we will focus on the low complexity uniform group size approach.

The goal is to minimize the MSE defined by
%%%%%%%%%%%%%%%%%%%%%%%%%%%%%%%%%%%%%%%%%%%%%%%%%%%%%%%%%%%%%%%%%%%%%%%%%%%%%%%%%%%%%%%%%%%%%%%%%%%%%%%%%%%%%%%%%%%
\begin{equation}
{\mathbb E}\{\|\bm h - \hat{\bm h}_{\rm b}\|^2\}={\mathbb E}\{(\hat{\bm
h}_{\rm b}-{\mathbb E}\{\hat{\bm h}_{\rm b}\})^{H}(\hat{\bm h}_{\rm b}-{\mathbb
E}\{\hat{\bm h}_{\rm b}\})\}+\|{\mathbb E}\{\hat{\bm h}_{\rm b}\}-\bm
h\|^{2}.
\label{equ:costfunc4vaopt}
\end{equation}Note that
%%%%%%%%%%%%%%%%%%%%%%%%%%%%%%%%%%%%%%%%%%%%%%%%%%%%%%%%%%%%%%%%%%%%%%%%%%%%%%%%%%%%%%%%%%%%%%%%%%%%%%%%%%%%%%%%%%%
\begin{equation}
\hat{\bm h}_{\rm b}-{\mathbb E}\{\hat{\bm h}_{\rm b}\}=(\hat{\bm H}_{\rm LS}-{\bm H})({\bf 1}_{S}+\boldsymbol \alpha)
\end{equation} and we have $\hat{\bm H}_{\rm LS}={\bm H}+\bm N$ where
%%%%%%%%%%%%%%%%%%%%%%%%%%%%%%%%%%%%%%%%%%%%%%%%%%%%%%%%%%%%%%%%%%%%%%%%%%%%%%%%%%%%%%%%%%%%%%%%%%%%%%%%%%%%%%%%%%%%
%\begin{equation}
%\bm N=\begin{bmatrix}
% \tilde{n}(1)                & 0   \\
% \vdots                             &  \vdots    \\
% \tilde{n}(\frac{L}{2})      & 0\\
% 0                                  & \tilde{n}(\frac{L}{2}+1)              \\
%  \vdots                            & \vdots      \\
% 0                                  &  \tilde{n}(L)\\
%\end{bmatrix}
%\end{equation}
%%%%%%%%%%%%%%%%%%%%%%%%%%%%%%%%%%%%%%%%%%%%%%%%%%%%%%%%%%%%%%%%%%%%%%%%%%%%%%%%%%%%%%%%%%%%%%%%%%%%%%%%%%%%%%%%%%%
\begin{equation}
\bm N=\begin{bmatrix}
 \tilde{\bm n}_{1}           &\bm 0              &\dots    & \bm 0                 \\
 \vdots                      & \tilde{\bm n}_{2}  &\ddots   & \vdots                         \\
 \vdots                      & \ddots            &\ddots   &  \vdots                        \\
 \bm 0                       & \dots             &\dots    &\tilde{\bm n}_{S}      \\
\end{bmatrix}
\end{equation} and %$ \tilde{\bm n}_{s}=[ \tilde{n}(s-1)L/S+1, \dots, \tilde{n}sL/S]$.
$\tilde{\bm n}=(\bm X^{H}\bm X)^{-1}\bm X^{H}\bm n=[\tilde{\bm n}_{1},\tilde{\bm n}_{2},\dots, \tilde{\bm n}_{S}]^{T}\in \bm C^{L\times 1}$, assuming that all the elements in this equivalent noise vector are independent and identically distributed (i.i.d.) random variables. Hence, we have
%%%%%%%%%%%%%%%%%%%%%%%%%%%%%%%%%%%%%%%%%%%%%%%%%%%%%%%%%%%%%%%%%%%%%%%%%%%%%%%%%%%%%%%%%%%%%%%%%%%%%%%%%%%%%%%%%%%
\begin{equation}
{\mathbb E}\{\big(\hat{\bm
h}_{\rm b}-{\mathbb E}\{\hat{\bm h}_{\rm b}\}\big)^{H}\big(\hat{\bm h}_{\rm b}-{\mathbb E}\{\hat{\bm h}_{\rm b}\}\big)\}=\tilde {\sigma}^{2}({\bf 1}_{S}+\boldsymbol \alpha)^{H}({\bf 1}_{S}+\boldsymbol \alpha)
\label{equ:varpartofMSE}
\end{equation} Note that $\tilde {\sigma}^{2}$ equals the variance of the equivalent noise $\tilde{\bm n}$ times the length of the group. We also have
%%%%%%%%%%%%%%%%%%%%%%%%%%%%%%%%%%%%%%%%%%%%%%%%%%%%%%%%%%%%%%%%%%%%%%%%%%%%%%%%%%%%%%%%%%%%%%%%%%%%%%%%%%%%%%%%%%%
\begin{equation}
\|{\mathbb E}\{\hat{\bm h}_{\rm b}\}-\bm h\|^{2}= {\boldsymbol \alpha}^{H}\bm H^{H}\bm H{\boldsymbol \alpha}
\label{equ:biasepartofMSE}
\end{equation}
Finally, the optimal solution of the vector ${\boldsymbol \alpha}$ that minimizes \eqref{equ:costfunc4vaopt} is given by
%%%%%%%%%%%%%%%%%%%%%%%%%%%%%%%%%%%%%%%%%%%%%%%%%%%%%%%%%%%%%%%%%%%%%%%%%%%%%%%%%%%%%%%%%%%%%%%%%%%%%%%%%%%%%%%%%%%
\begin{equation}
\boldsymbol \alpha_{\rm opt} = - \tilde {\sigma}^{2} \big(\tilde {\sigma}^{2}\bm I_{S}+\bm H^{H}\bm H\big)^{-1}{\bf 1}_{S}.
\label{equ:aopt}
\end{equation} and we have
%%%%%%%%%%%%%%%%%%%%%%%%%%%%%%%%%%%%%%%%%%%%%%%%%%%%%%%%%%%%%%%%%%%%%%%%%%%%%%%%%%%%%%%%%%%%%%%%%%%%%%%%%%%%%%%%%%%
\begin{equation}
\hat{\bm h}_{\rm b, opt}=\hat{\bm H}_{\rm LS}({\bf 1}_{S}+\boldsymbol \alpha_{\rm opt})
\end{equation}

Note that this equation is a general expression for different numbers of groups. The complexity of this algorithm is very low because the inverse matrix required to calculate the $\boldsymbol \alpha_{\rm opt}$ is a diagonal matrix. Hence, this estimator combined with the conventional RLS algorithm will only introduce an additional complexity that is linear in the length of the parameter vector $L$ and the number of groups $S$. If the group size equals $L$, then the GSE converges to the biased estimator with only one shrinkage factor. In the following section,
adaptive algorithms will be developed to compute the best GSE with a given group size.

\subsection{Proposed GSE: Adaptive Algorithms}

It should be noted that the optimal solution of the biased estimator requires some prior knowledge of the system, which is the matrix $\bm H^{H}\bm H$ and the scalar term $\tilde {\sigma}^{2}$ for calculation of the vector ${\boldsymbol \alpha}$. In addition, the LS channel estimator is also required. The LS channel estimator can be recursively calculated by the RLS adaptive algorithm that is detailed in Section \ref{sec:systemmodelsce}. In this work, we propose LMS-based adaptive algorithms that enable us to estimate the vector ${\boldsymbol \alpha}$ without prior knowledge of the channel and the noise variance. Substituting \eqref{equ:biasepartofMSE} and \eqref{equ:varpartofMSE} into \eqref{equ:costfunc4vaopt} and considering the MSE cost function as a function of ${\boldsymbol \alpha}$, we can obtain a new cost function
%%%%%%%%%%%%%%%%%%%%%%%%%%%%%%%%%%%%%%%%%%%%%%%%%%%%%%%%%%%%%%%%%%%%%%%%%%%%%%%%%%%%%%%%%%%%%%%%%%%%%%%%%%%%%%%%%%%
\begin{equation}
f(\boldsymbol \alpha)=\tilde {\sigma}^{2}({\bf 1}_{S}+\boldsymbol \alpha)^{H}({\bf 1}_{S}+\boldsymbol \alpha)+{\boldsymbol \alpha}^{H}\bm H^{H}\bm H{\boldsymbol \alpha}.
\end{equation} The gradient of $f(\boldsymbol \alpha)$ with respect to ${\boldsymbol \alpha}$ is given by
%%%%%%%%%%%%%%%%%%%%%%%%%%%%%%%%%%%%%%%%%%%%%%%%%%%%%%%%%%%%%%%%%%%%%%%%%%%%%%%%%%%%%%%%%%%%%%%%%%%%%%%%%%%%%%%%%%%
\begin{equation}
\bm g_{\boldsymbol \alpha}=\tilde {\sigma}^{2}({\bf 1}_{S}+\boldsymbol \alpha)+\bm H^{H}\bm H{\boldsymbol \alpha}.
\end{equation} Note that, because $\boldsymbol \alpha$ is a real-valued vector, there is a factor of 2 for this gradient. In what follows, this factor is absorbed into the step size of gradient-type recursions. Hence, the LMS-based update equation of the vector $\boldsymbol \alpha$ for the $(i+1)$-th time slot can be expressed as

%%%%%%%%%%%%%%%%%%%%%%%%%%%%%%%%%%%%%%%%%%%%%%%%%%%%%%%%%%%%%%%%%%%%%%%%%%%%%%%%%%%%%%%%%%%%%%%%%%%%%%%%%%%%%%%%%%
\begin{equation}
\hat{\boldsymbol \alpha} (i+1)=\hat{\boldsymbol \alpha} (i)-\mu\hat{\bm g}_{\boldsymbol \alpha}(i),
\end{equation} where $\mu$ is the step size of the LMS algorithm and the estimated gradient vector is given by
%%%%%%%%%%%%%%%%%%%%%%%%%%%%%%%%%%%%%%%%%%%%%%%%%%%%%%%%%%%%%%%%%%%%%%%%%%%%%%%%%%%%%%%%%%%%%%%%%%%%%%%%%%%%%%%%%%%
\begin{equation}
\hat{\bm g}_{\boldsymbol \alpha}(i)=\hat{\tilde {\sigma}}^{2}(i)\big({\bf 1}_{S}+\hat{\boldsymbol \alpha} (i)\big)+\hat{\bm P}_{m}(i)\hat{\boldsymbol \alpha} (i).
\end{equation}

Here, $\hat{\tilde {\sigma}}^{2}(i)$ is the estimated equivalent
noise variance and the diagonal matrix $\hat {\bm P}_{m}(i)$ is
defined as the estimator of the matrix  $\bm H^{H}\bm H$, the main
diagonal vector of this matrix is defined as ${\rm diag}[\hat {\bm
P}_{m}(i)]=[\hat { P}_{m,1}(i),\dots,\hat { P}_{m,S}(i)]$ . In this
work, we adopt the instantaneous estimator as $\hat{\tilde
{\sigma}}^{2}(i)=\|\hat{\bm h}_{\rm RLS}(i)-\hat{\bm h}(i)\|^{2}/S$,
where $\hat{\bm h}_{\rm RLS}(i)$ is the RLS channel estimator and
$\hat{\bm h}(i)=\frac{1}{i}\sum^{i}_{j=1}\hat{\bm h}_{\rm RLS}(j)$
represent the time averaged channel estimator. Note that $\bm
H^{H}\bm H\in \mathbb C^{S\times S}$ is a diagonal matrix with its
$s$-th diagonal element equals
$h_{\Sigma,s}=\sum^{sL/S}_{l=(s-1)L/S+1} |h(l)|^2$.
%%%%%%%%%%%%%%%%%%%%%%%%%%%%%%%%%%%%%%%%%%%%%%%%%%%%%%%%%%%%%%%%%%%%%%%%%%%%%%%%%%%%%%%%%%%%%%%%%%%%%%%%%%%%%%%%%%%%
%\begin{equation}
%\bm H^{H}\bm H=\begin{bmatrix}
% \sum^{L/S}_{l=1} |h(l)|^2        & 0                                 & \dots    & 0\\
% 0                                & \sum^{2L/S}_{l=L/S+1} |h(l)|^2    & \dots    &\vdots\\
% \vdots                           & \vdots                            & \dots    &\vdots\\
% \vdots                           & \vdots                            & \cdots   &\vdots\\
%  0                                & \dots                            & 0        & \sum^{L}_{l=(S-1)L/S+1} |h(l)|^2\\
%\end{bmatrix}.
%\label{equ:HhHexpression}
%\end{equation} To simplify the following derivation, we define scalar terms $h_{\Sigma,s}=\sum^{sL/S}_{l=(s-1)L/S+1} |h(l)|^2$.

Hence, the elements in the optimal solution can be expressed as
$\alpha_{\rm opt,s}=-(1+h_{\Sigma,s}/\tilde {\sigma}^{2})^{-1}$. If
we use the matrix $\hat {\bm P}_{m}(i)$ to replace the matrix $\bm
H^{H}\bm H$, the estimated optimal solution becomes
%%%%%%%%%%%%%%%%%%%%%%%%%%%%%%%%%%%%%%%%%%%%%%%%%%%%%%%%%%%%%%%%%%%%%%%%%%%%%%%%%%%%%%%%%%%%%%%%%%%%%%%%%%%%%%%%%%%
\begin{equation}
\hat{\alpha}_{\rm opt,s}=-\frac{\tilde {\sigma}^{2}}{\tilde {\sigma}^{2}+\hat { P}_{m,s}(i)}.
\end{equation} Recall that we assume that the shrinkage factors for each group are larger than zero but smaller than one. It can be found that if $\hat {P}_{m,s}(i)< 0$, this assumption no longer holds. In addition, if $\hat {P}_{m,s}(i)\rightarrow \infty$, the shrinkage factor converges to zero, and the biased estimator actually converges to the unbiased estimator. So we should constrain the values of $\hat {P}_{m,s}(i)$ into the range $(0,+\infty)$.

In order to determine the diagonal matrices $\hat {\bm P}_{m}(i)$ for each time instant, two approaches are developed in this work. In the first approach, which is named estimator based (GSE-EB) method, the matrices $\hat {\bm P}_{m}(i)$ are replaced by the diagonal matrices $\hat{\bm H}_{\rm RLS}^{H}(i)\hat{\bm H}_{\rm RLS}(i)$, where $\hat{\bm H}_{\rm RLS}(i)$ is the RLS estimator of ${\bm H}$. Note that, when the number of groups is only one, the GSE-EB method will lead to an optimal shrinkage factor $\alpha$ that has the same expression as the SBME that is proposed in \cite{zbenhaim2007}. However, the knowledge of the noise variance is not required in our work. In the second approach, which is named automatic tuning (GSE-AT) method, an LMS-based algorithm is proposed to update the diagonal matrices $\hat {\bm P}_{m}(i)$.

For the GSE-EB method, the estimation of $\hat {\bm P}_{m}(i)$ is only determined by the RLS estimator. If the effective spreading codes and the channel information is known, the RLS algorithm can be initialized efficiently. Here, we consider a general scenario where all these
quantities are unknown and the initialization of the RLS algorithm is an all zero vector, which means the beginning stage of the RLS algorithm is
not very accurate. In order to improve the convergence rate of the proposed GSE schemes, we develop the following GSE-AT algorithm.
For each time instant, $\hat {\bm P}_{m}(i)$ is firstly set to $\hat{\bm H}_{\rm RLS}^{H}(i)\hat{\bm H}_{\rm RLS}(i)$ as in the GSE-EB algorithm, then we consider $\hat {P}_{m,s}(i)$ as the variables of the MSE cost function, where $\hat {P}_{m,s}(i)$ are the diagonal elements of the diagonal matrix $\hat {\bm P}_{m}(i)$ and $s=1,2,\dots,S$. Then we develop an LMS adaptive recursion to further adapt these values and improve the estimation accuracy for each time instant. Let us reexpress the MSE cost function as shown in \eqref{equ:costfunc4vaopt} as follows. Here, we omit the time index $i$ for simplicity
%%%%%%%%%%%%%%%%%%%%%%%%%%%%%%%%%%%%%%%%%%%%%%%%%%%%%%%%%%%%%%%%%%%%%%%%%%%%%%%%%%%%%%%%%%%%%%%%%%%%%%%%%%%%%%%%%%%
\begin{equation}
f(\hat {\bm P}_{m})=\tilde {\sigma}^{2}\big({\bf 1}_{S}+\boldsymbol \alpha\big)^{H}\big({\bf 1}_{S}+\boldsymbol \alpha\big)+{\boldsymbol \alpha}^{H}\hat {\bm P}_{m}{\boldsymbol \alpha},
\end{equation}where $\boldsymbol \alpha=- \tilde {\sigma}^{2} \big(\tilde {\sigma}^{2}\bm I_{S}+\hat {\bm P}_{m}\big)^{-1}{\bf 1}_{S}$ is also a function of $\hat {\bm P}_{m}$. Expanding this cost function, we have
%%%%%%%%%%%%%%%%%%%%%%%%%%%%%%%%%%%%%%%%%%%%%%%%%%%%%%%%%%%%%%%%%%%%%%%%%%%%%%%%%%%%%%%%%%%%%%%%%%%%%%%%%%%%%%%%%%%
\begin{equation}
f([\hat { P}_{m,1},\dots,\hat { P}_{m,S}])=\tilde {\sigma}^{2}\sum^{S}_{s=1}\big(1+\alpha_{s}\big)^{2}+\sum^{S}_{s=1}\alpha_{s}^2\hat { P}_{m,s}.
\label{equ:costfunc4AT1}
\end{equation}Hence, for each group, the corresponding $\hat {P}_{m,s}$ can be obtained by the following equation
%%%%%%%%%%%%%%%%%%%%%%%%%%%%%%%%%%%%%%%%%%%%%%%%%%%%%%%%%%%%%%%%%%%%%%%%%%%%%%%%%%%%%%%%%%%%%%%%%%%%%%%%%%%%%%%%%%
\begin{equation}
\hat {P}_{m,s}(c+1)=\hat {P}_{m,s}(c)-\mu_{p}\hat{g}_{s},
\label{equ:hatpmiat1_6}
\end{equation}where $c$ is the iteration index and $\hat{g}_{s}$ is the estimator of the gradient of the function \eqref{equ:costfunc4AT1} with respect to $\hat {P}_{m,s}$, which is given by
%%%%%%%%%%%%%%%%%%%%%%%%%%%%%%%%%%%%%%%%%%%%%%%%%%%%%%%%%%%%%%%%%%%%%%%%%%%%%%%%%%%%%%%%%%%%%%%%%%%%%%%%%%%%%%%%%%%
\begin{equation}
\hat{g}_{s}=2\hat{\tilde {\sigma}}^{2}\big(1+\hat{\alpha}_{s}\big)\frac{\hat{\tilde {\sigma}}^{2}}{\big(\hat{\tilde {\sigma}}^{2}+\hat {P}_{m,s}\big)^{2}}+2\hat{\alpha}_{s}\hat {P}_{m,s}\frac{\hat{\tilde {\sigma}}^{2}}{\big(\hat{\tilde {\sigma}}^{2}+\hat {P}_{m,s}\big)^{2}}+\hat{\alpha}_{s}^{2}.
\label{equ:garident4atsce}
\end{equation}

In order to obtain a low complexity solution, for the GSE-AT algorithm, we set the iteration index $c$ to 1, which means for each time instant we only update the values of $\hat {P}_{m,s}$ once. As pointed out previously, the values of $\hat {\bm P}_{m}(i)$ should be constrained in the range $(0,+\infty)$. Hence, for the GSE-AT algorithm, if the updated values are negative, then we set these values to the same values of GSE-EB algorithm. In Table \ref{tab:Biasedfors1_6}, the proposed biased estimators using two approaches to calculate $\hat{P}_{\rm m}(i)$ are summarized.

Note that the GSE-EB approach (where $\hat{P}_{\rm m}(i)=\hat{\bm H}_{\rm RLS}^{H}(i)\hat{\bm H}_{\rm RLS}(i)$) requires $2L+3S+2$ complex multiplications and $3L+S$ complex additions for one update of the shrinkage factors. For the GSE-AT approach, in which $\hat{P}_{\rm m}(i)$ is updated by using equation \eqref{equ:hatpmiat1_6}, the number of complex multiplications required to update the shrinkage factor is $2L+13S+2$ and the number of complex additions required is $3L+7S$. It will be demonstrated by the simulations that the performance of the GSE-AT approach is better than the GSE-EB approach, while the GSE-EB approach has a lower complexity.

\begin{table*}
\centering \caption{\normalsize GSE for SCE-RLS in SC-FDE DS-UWB
Systems}
\begin{tabular}{|l|l|}
\hline
\large {Proposed GSE-EB} & \large {Proposed GSE-AT}\rule{0pt}{2.6ex}\\
\hline
\bfseries {1. Initialization:} & \bfseries {1. Initialization:}  \rule{0pt}{2.6ex}\\
$\hat{\boldsymbol \alpha}(1)=\bm 0\in \mathbb C^{S\times 1}$                                                                                       &$\hat{\boldsymbol \alpha}(1)=\bm 0\in \mathbb C^{S\times 1}$  \\
Set value of $\mu$                                                                                  &Set values of $\mu$ and $\mu_{p}$\\
\hline
\bfseries {2. Calculate the biased estimator:} & \bfseries {2. Calculate the biased estimator:} \rule{0pt}{2.6ex} \\
For $i=1,2,\dots$                                                                                  &For $i=1,2,\dots$\rule{0pt}{2.6ex}\\
$\hat{\bm h}_{\rm b}(i)=\hat{\bm H}_{\rm RLS}(i)({\bf 1}_{S}+\hat{\boldsymbol \alpha}(i))$                           &$\hat{\bm h}_{\rm b}(i)=\hat{\bm H}_{\rm RLS}(i)({\bf 1}_{S}+\hat{\boldsymbol \alpha}(i))$ \rule{0pt}{2.6ex}\\
\hline
\bfseries {3. Calculate the shrinkage factor:} & \bfseries {3. Calculate the shrinkage factor:} \rule{0pt}{2.6ex} \\
$\hat{\bm h}(i)=\frac{1}{i}\sum^{i}_{j=1}\hat{\bm h}_{\rm RLS}(j)$                              &$\hat{\bm h}(i)=\frac{1}{i}\sum^{i}_{j=1}\hat{\bm h}_{\rm RLS}(j)$\rule{0pt}{2.6ex}\\
$\hat{\tilde {\sigma}}^{2}(i)=\|\hat{\bm h}_{\rm RLS}(i)-\hat{\bm h}(i)\|^{2}/S$ &$\hat{\tilde {\sigma}}^{2}(i)=\|\hat{\bm h}_{\rm RLS}(i)-\hat{\bm h}(i)\|^{2}/S$\rule{0pt}{2.6ex}\\
$\hat {\bm P}_{m}(i)=\hat{\bm H}_{\rm RLS}^{H}(i)\hat{\bm H}_{\rm RLS}(i)$                                      &$\hat {\bm P}_{m}(i)=\hat{\bm H}_{\rm RLS}^{H}(i)\hat{\bm H}_{\rm RLS}(i)$ \rule{0pt}{2.6ex}\\
                                                                                                         &For $s=1,2,\dots,S$\rule{0pt}{2.6ex}\\
$\hat{\bm g}_{\boldsymbol \alpha}(i)=\hat{\tilde {\sigma}}^{2}(i)({\bf 1}_{S}+\hat{\boldsymbol \alpha} (i))+\hat{\bm P}_{m}(i)\hat{\boldsymbol \alpha} (i),$                    &$\hat {P}_{m,s}(i)=\hat {P}_{m,s}(i)-\mu_{p}\hat{g}_{s}(i)$, where $\hat{g}_{s}(i)$ is given in \eqref{equ:garident4atsce}.\rule{0pt}{2.6ex}\\
                                                                                                         &  If $\hat {P}_{m,s}(i)<0$, set $\hat {\bm P}_{m}(i)=\hat{\bm H}_{\rm RLS}^{H}(i)\hat{\bm H}_{\rm RLS}(i)$ break;\\
                                                                                                         &End For.\rule{0pt}{2.6ex}\\
$\hat{\boldsymbol \alpha} (i+1)=\hat{\boldsymbol \alpha} (i)-\mu\hat{\bm g}_{\boldsymbol \alpha}(i).$                                                       &$\hat{\bm g}_{\boldsymbol \alpha}(i)=\hat{\tilde {\sigma}}^{2}(i)({\bf 1}_{S}+\hat{\boldsymbol \alpha} (i))+\hat{\bm P}_{m}(i)\hat{\boldsymbol \alpha} (i),$\\
                                                                                                       &$\hat{\boldsymbol \alpha} (i+1)=\hat{\boldsymbol \alpha} (i)-\mu\hat{\bm g}_{\boldsymbol \alpha}(i).$\\
\hline
\end{tabular}
\label{tab:Biasedfors1_6}
\end{table*}

\section{Proposed GSE for interference suppression scenario}
\label{sec:IS}

%%%%%%%%%%%%%%%%%%%%%%%%%%%%%%%%%%%%%%%%%%%%%%%%%%%%%%%%%%%%%%%%%%%%%%%%%%%%%%
\subsection{Proposed GSE: Optimal Solution}
For the interference suppression scenario, the biased estimator with a shrinkage factor is given by \cite{myEW2011}
%%%%%%%%%%%%%%%%%%%%%%%%%%%%%%%%%%%%%%%%%%%%%%%%%%%%%%%%%%%%%%%%%%%%%%%%%%%%%%
\begin{equation}
\hat{\bm w}_{\rm b}=(1+\alpha)\hat{\bm w}_{\rm LS}.
\end{equation}where $\hat{\bm w}_{\rm LS}\in \mathbb C^{M\times 1}$ is the LS estimator of the receive filter. The proposed GSE can be expressed as follows:
%%%%%%%%%%%%%%%%%%%%%%%%%%%%%%%%%%%%%%%%%%%%%%%%%%%%%%%%%%%%%%%%%%%%%%%%%%%%%%%%%%%%%%%%%%%%%%%%%%%%%%%%%%%%%%%%%%%
\begin{equation}
%\begin{split}
\hat{\bm w}_{\rm b} =\begin{bmatrix}
 (1+\alpha_{w,1})\hat{w}_{\rm LS}(1)    \\
 \vdots      \\
(1+\alpha_{w,1})\hat{w}_{\rm LS}(\frac{L}{S})\\
 (1+\alpha_{w,2})\hat{w}_{\rm LS}(\frac{L}{S}+1)              \\
  \vdots      \\
  (1+\alpha_{w,2})\hat{w}_{\rm LS}(\frac{2L}{S})\\
   \vdots      \\
    (1+\alpha_{w,S})\hat{w}_{\rm LS}(\frac{(S-1)L}{S}+1)              \\
  \vdots      \\
  (1+\alpha_{w,S})\hat{w}_{\rm LS}(L)\\
\end{bmatrix}
= \hat{\bm W}_{\rm LS} ({\bf 1}_{S}+\boldsymbol \alpha_{w})%\left(\begin{bmatrix}
%  1    \\
%  1    \\
%\end{bmatrix}+\begin{bmatrix}
%  \alpha_{1}    \\
%  \alpha_{2}    \\
%\end{bmatrix}\right)
%\end{split}
\label{equ:proposedwb}
\end{equation}where $\hat{\bm W}_{\rm LS}\in \mathbb C^{M\times S}$ is a block diagonal matrix that is constructed by the elements from $\hat{\bm w}_{\rm LS}$ and zeros, $S$ is the number of groups. Moreover, we define the $S$-dimensional column vectors ${\bf 1}_{S} =[1, 1,\dots, 1]^{T} $ and $\boldsymbol \alpha_{w} =[\alpha_{w,1} , \alpha_{w,2},\dots, \alpha_{w,S} ]^{T} $. The quantities $\alpha_{w,s}$ are real-valued variables and $(1+\alpha_{w,s})$ is defined as the shrinkage factor for the $s$-th group of coefficients that is larger than zero but smaller than one, where $s=1,2,\dots,S$.

The objective of the GSE is to achieve a smaller MSE than the LS algorithm, which can be expressed as
%%%%%%%%%%%%%%%%%%%%%%%%%%%%%%%%%%%%%%%%%%%%%%%%%%%%%%%%%%%%%%%%%%%%%%%%%%%%%%%%%%%%%%%%%%%%%%%%%%%%%%%%%%%%%%%%%%%
\begin{equation}
{\mathbb E}[\|\bm w_{\rm o} - \hat{\bm w}_{\rm b}\|^2]\leq{\mathbb E}[\|\bm
w_{\rm o} - \hat{\bm w}_{\rm LS}\|^2]. \label{equ:targetofbiased2}
\end{equation}
It should be noted that the problem that we want to solve for both parameter estimation and interference suppression scenarios has a similar form.
Hence, we can follow the derivation as shown in Section \ref{sec:Optgse_sce}, and the optimal solution of the parameter vector $\boldsymbol \alpha_{w}$ is given by
%%%%%%%%%%%%%%%%%%%%%%%%%%%%%%%%%%%%%%%%%%%%%%%%%%%%%%%%%%%%%%%%%%%%%%%%%%%%%%%%%%%%%%%%%%%%%%%%%%%%%%%%%%%%%%%%%%%
\begin{equation}
\boldsymbol \alpha_{w,\rm opt} = - \tilde {\sigma}_{w}^{2} \big(\tilde {\sigma}_{w}^{2}\bm I_{S}+\bm W^{H}\bm W\big)^{-1}{\bf 1}_{S}.
\label{equ:aopt_w}
\end{equation} where $\bm W=\mathbb E [\hat{\bm W}_{\rm LS}]$, $\tilde {\sigma}_{w}^{2}=v_{w}/S$ and we have
%%%%%%%%%%%%%%%%%%%%%%%%%%%%%%%%%%%%%%%%%%%%%%%%%%%%%%%%%%%%%%%%%%%%%%%%%%%%%%%%%%%%%%%%%%%%%%%%%%%%%%%%%%%%%%%%%%%
\begin{equation}
\hat{\bm w}_{\rm b,opt}=\hat{\bm W}_{\rm LS}({\bf 1}_{S}+\boldsymbol \alpha_{w,\rm opt})
\end{equation}

%%%%%%%%%%%%%%%%%%%%%%%%%%%%%%%%%%%%%%%%%%%%%%%%%%%%%%%%%%%%%%%%%%%%%%%%%%%%%%
\subsection{Proposed GSE: Adaptive Algorithms}
In this section, the LMS-based adaptive algorithms are proposed to estimate the vector ${\boldsymbol \alpha_{w}}$. First, we consider the MSE cost function as a function of ${\boldsymbol \alpha_{w}}$, i.e.,
%%%%%%%%%%%%%%%%%%%%%%%%%%%%%%%%%%%%%%%%%%%%%%%%%%%%%%%%%%%%%%%%%%%%%%%%%%%%%%%%%%%%%%%%%%%%%%%%%%%%%%%%%%%%%%%%%%%
\begin{equation}
f(\boldsymbol \alpha_{w})=\tilde {\sigma}_{w}^{2}({\bf 1}_{S}+\boldsymbol \alpha_{w})^{H}({\bf 1}_{S}+\boldsymbol \alpha_{w})+{\boldsymbol \alpha_{w}}^{H}\bm W^{H}\bm W{\boldsymbol \alpha_{w}}.
\end{equation} The gradient of $f(\boldsymbol \alpha_{w})$ with respect to ${\boldsymbol \alpha_{w}^{*}}$ is given by
$\bm g_{w}=\tilde {\sigma}_{w}^{2}({\bf 1}_{S}+\boldsymbol \alpha_{w})+\bm W^{H}\bm W{\boldsymbol \alpha_{w}}.$ Hence, the LMS-based update equation of the vector $\boldsymbol \alpha_{w}$ for the $(i+1)$-th time slot can be expressed as

%%%%%%%%%%%%%%%%%%%%%%%%%%%%%%%%%%%%%%%%%%%%%%%%%%%%%%%%%%%%%%%%%%%%%%%%%%%%%%%%%%%%%%%%%%%%%%%%%%%%%%%%%%%%%%%%%%%
\begin{equation}
\hat{\boldsymbol \alpha}_{w} (i+1)=\hat{\boldsymbol \alpha}_{w} (i)-\mu_{w}\hat{\bm g}_{w}(i),
\end{equation} where $\mu_{w}$ is the step size of the LMS algorithm and the estimated gradient vector is given by
%%%%%%%%%%%%%%%%%%%%%%%%%%%%%%%%%%%%%%%%%%%%%%%%%%%%%%%%%%%%%%%%%%%%%%%%%%%%%%%%%%%%%%%%%%%%%%%%%%%%%%%%%%%%%%%%%%%%
\begin{equation}
\hat{\bm g}_{w}(i)=
\hat{\tilde {\sigma}}_{w}^{2}(i)({\bf 1}_{S}+\hat{\boldsymbol \alpha}_{w} (i))+\hat{\bm P}_{w}(i)\hat{\boldsymbol \alpha}_{w} (i),
\end{equation}where $\hat{\tilde {\sigma}}_{w}^{2}(i)$ is the estimated equivalent noise variance and the diagonal matrix $\hat {\bm P}_{w}(i)\in \mathbb C^{S\times S}$ is defined as the estimator of the matrix  $\bm W^{H}\bm W$,
the main diagonal vector of this matrix is defined as ${\rm diag}[\hat {\bm P}_{w}(i)]=[\hat { P}_{w,1}(i),\dots,\hat { P}_{w,S}(i)]$.
The instantaneous estimator of $\hat{\tilde {\sigma}}_{w}^{2}(i)$ is given by
%%%%%%%%%%%%%%%%%%%%%%%%%%%%%%%%%%%%%%%%%%%%%%%%%%%%%%%%%%%%%%%%%%%%%%%%%%%%%%%%%%%%%%%%%%%%%%%%%%%%%%%%%%%%%%%%%%%%
%\begin{equation}
$\hat{\tilde {\sigma}}_{w}^{2}(i)=\big(\hat{\bm w}_{\rm RLS}(i)-\bm w_{\rm
o}(i)\big)^{H}\big(\hat{\bm w}_{\rm RLS}(i)-\bm w_{\rm o}(i)\big),$
%\label{equ:varw_6}
%\end{equation}
where $\bm w_{\rm o}$ is replaced by the time averaged RLS estimator, that is $\bm w_{\rm o}(i)=\frac{1}{i}\sum^{i}_{j=1}\hat{\bm w}_{\rm RLS}(j)$.

In order to determine the diagonal matrices $\hat {\bm P}_{w}(i)$ for each time instant, the GSE-EB method and the GSE-AT method are developed in the interference suppression scenario. In the GSE-EB approach, the matrices $\hat {\bm P}_{w}(i)$ are replaced by the diagonal matrices $\hat{\bm W}_{\rm RLS}^{H}(i)\hat{\bm W}_{\rm RLS}(i)$. However, because we assume that the initialization of the RLS algorithm is an all zero vector, the beginning stage of the RLS algorithm is not very accurate. Hence, in order to improve the convergence rate of the proposed GSE schemes, we develop the GSE-AT algorithm. For each time instant, the matrix $\hat {\bm P}_{w}(i)$ is firstly set to $\hat{\bm W}_{\rm RLS}^{H}(i)\hat{\bm W}_{\rm RLS}(i)$ as in the GSE-EB algorithm, then we consider $\hat {P}_{w,s}(i)$ as the variables of the MSE cost function, where $\hat {P}_{w,s}(i)$ are the diagonal elements of the diagonal matrix $\hat {\bm P}_{w}(i)$ and $s=1,2,\dots,S$. Then we develop an LMS adaptive equation to further adapt these values and improve the estimation accuracy for each time instant. Here, we omit the time index $i$ for simplicity and have
%%%%%%%%%%%%%%%%%%%%%%%%%%%%%%%%%%%%%%%%%%%%%%%%%%%%%%%%%%%%%%%%%%%%%%%%%%%%%%%%%%%%%%%%%%%%%%%%%%%%%%%%%%%%%%%%%%
\begin{equation}
f([\hat { P}_{w,1},\dots,\hat { P}_{w,S}])=\tilde {\sigma}_{w}^{2}\sum^{S}_{s=1}\big(1+\alpha_{w,s}\big)^{2}+\sum^{S}_{s=1}\alpha_{w,s}^2\hat { P}_{w,s}.
\label{equ:costfunc4ATis}
\end{equation}Hence, for each group, the corresponding $\hat {P}_{w,s}$ can be updated by the following equation
%%%%%%%%%%%%%%%%%%%%%%%%%%%%%%%%%%%%%%%%%%%%%%%%%%%%%%%%%%%%%%%%%%%%%%%%%%%%%%%%%%%%%%%%%%%%%%%%%%%%%%%%%%%%%%%%%%
\begin{equation}
\hat {P}_{w,s}(c+1)=\hat {P}_{w,s}(c)-\mu_{s}\hat{g}_{w,s},
\label{equ:hatpmiat1_62}
\end{equation}where $c$ is the iteration index and $\hat{g}_{w,s}$ is the estimator of the gradient of the cost function with respect to $\hat {P}_{m,s}$, which is given by
%%%%%%%%%%%%%%%%%%%%%%%%%%%%%%%%%%%%%%%%%%%%%%%%%%%%%%%%%%%%%%%%%%%%%%%%%%%%%%%%%%%%%%%%%%%%%%%%%%%%%%%%%%%%%%%%%%%
\begin{equation}
\hat{g}_{w,s}=2\hat{\tilde {\sigma}}^{2}\big(1+\hat{\alpha}_{s}\big)\frac{\hat{\tilde {\sigma}}^{2}}{\big(\hat{\tilde {\sigma}}^{2}+\hat {P}_{m,s}\big)^{2}}+2\hat{\alpha}_{s}\hat {P}_{m,s}\frac{\hat{\tilde {\sigma}}^{2}}{\big(\hat{\tilde {\sigma}}^{2}+\hat {P}_{m,s}\big)^{2}}+\hat{\alpha}_{s}^{2}.
\label{equ:garident4atIS}
\end{equation}

In order to obtain a low complexity solution, we set the iteration index $c$ to 1 for the GSE-AT algorithm. Note that if the updated values in the GSE-AT algorithm become negative, then we set these values to the same as obtained in the GSE-EB algorithm. In Table \ref{tab:Biasedfors1_62}, the proposed biased estimators with two approaches to calculate $\hat{P}_{\rm w}(i)$ are summarized.

Note that the GSE-EB approach (where $\hat{P}_{\rm w}(i)=\hat{\bm W}_{\rm RLS}^{H}(i)\hat{\bm W}_{\rm RLS}(i)$), requires $2M+3S+2$ complex multiplications and $3M+S$ complex additions for one update of the shrinkage factors. For the GSE-AT approach, in which $\hat{P}_{\rm w}(i)$ is updated by using equation \eqref{equ:hatpmiat1_62}, the number of complex multiplications required to update the shrinkage factor is $2M+13S+2$ and the number of complex additions required is $3M+7S$. It will be demonstrated by the simulations that the performance of the GSE-AT approach is better than the GSE-EB approach, while the GSE-EB approach has a lower complexity.

\begin{table*}
\centering \caption{\normalsize GSE for Frequency domain receiver in SC-FDE DS-UWB
Systems}
\begin{tabular}{|l|l|}
\hline
\large {Proposed GSE-EB} & \large {Proposed GSE-AT}\\
\hline
\bfseries {1. Initialization:} & \bfseries {1. Initialization:} \rule{0pt}{2.6ex} \\
$\hat{\boldsymbol \alpha}_{w}(1)=\bm 0\in \mathbb C^{S\times 1}$                                                                                       &$\hat{\boldsymbol \alpha}_{w}(1)=\bm 0\in \mathbb C^{S\times 1}$  \rule{0pt}{2.6ex}\\
Set value of $\mu_{w}$                                                                                  &Set values of $\mu_{w}$ and $\mu_{s}$\\
\hline
\bfseries {2. Calculate the biased estimator:} & \bfseries {2. Calculate the biased estimator:} \rule{0pt}{2.6ex} \\
For $i=1,2,\dots$                                                                                  &For $i=1,2,\dots$\rule{0pt}{2.6ex}\\
$\hat{\bm w}_{\rm b}(i)=\hat{\bm W}_{\rm RLS}(i)({\bf 1}_{S}+\hat{\boldsymbol \alpha}_{w}(i))$                           &$\hat{\bm w}_{\rm b}(i)=\hat{\bm W}_{\rm RLS}(i)({\bf 1}_{S}+\hat{\boldsymbol \alpha}_{w}(i))$ \rule{0pt}{2.6ex}\\
\hline
\bfseries {3. Calculate the shrinkage factor:} & \bfseries {3. Calculate the shrinkage factor:} \rule{0pt}{2.6ex} \\
$\hat{\bm w}(i)=\frac{1}{i}\sum^{i}_{j=1}\hat{\bm w}_{\rm RLS}(j)$                              &$\hat{\bm w}(i)=\frac{1}{i}\sum^{i}_{j=1}\hat{\bm w}_{\rm RLS}(j)$\rule{0pt}{2.6ex}\\
$\hat{\tilde {\sigma}}_{w}^{2}(i)=\|\hat{\bm w}_{\rm RLS}(i)-\hat{\bm w}(i)\|^{2}/S$ &$\hat{\tilde {\sigma}}_{w}^{2}(i)=\|\hat{\bm w}_{\rm RLS}(i)-\hat{\bm w}(i)\|^{2}/S$\rule{0pt}{2.6ex}\\
$\hat {\bm P}_{w}(i)=\hat{\bm W}_{\rm RLS}^{H}(i)\hat{\bm W}_{\rm RLS}(i)$                                      &$\hat {\bm P}_{w}(i)=\hat{\bm W}_{\rm RLS}^{H}(i)\hat{\bm W}_{\rm RLS}(i)$ \rule{0pt}{2.6ex}\\
                                                                                                         &For $s=1,2,\dots,S$\rule{0pt}{2.6ex}\\
$\hat{\bm g}_{w}(i)=\hat{\tilde {\sigma}}_{w}^{2}(i)({\bf 1}_{S}+\hat{\boldsymbol \alpha}_{w} (i))+\hat{\bm P}_{w}(i)\hat{\boldsymbol \alpha}_{w} (i),$                    &$\hat {P}_{w,s}(i)=\hat {P}_{w,s}(i)-\mu_{s}\hat{g}_{w,s}(i)$, where $\hat{g}_{w,s}(i)$ is given in \eqref{equ:garident4atIS}.\rule{0pt}{2.6ex}\\
                                                                                                         &  If $\hat {P}_{w,s}(i)<0$, set $\hat {\bm P}_{w}(i)=\hat{\bm W}_{\rm RLS}^{H}(i)\hat{\bm W}_{\rm RLS}(i)$ break;\\
                                                                                                         &End For.\rule{0pt}{2.6ex}\\
$\hat{\boldsymbol \alpha}_{w} (i+1)=\hat{\boldsymbol \alpha}_{w} (i)-\mu_{w}\hat{\bm g}_{w}(i).$                                                       &$\hat{\bm g}_{w}(i)=\hat{\tilde {\sigma}}_{w}^{2}(i)({\bf 1}_{S}+\hat{\boldsymbol \alpha}_{w} (i))+\hat{\bm P}_{w}(i)\hat{\boldsymbol \alpha}_{w} (i),$\\
                                                                                                       &$\hat{\boldsymbol \alpha}_{w} (i+1)=\hat{\boldsymbol \alpha}_{w} (i)-\mu_{w}\hat{\bm g}_{w}(i).$\\
\hline
\end{tabular}
\label{tab:Biasedfors1_62}
\end{table*}

%%%%%%%%%%%%%%%%%%%%%%%%%%%%%%%%%%%%%%%%%%%%%%%%%%%%%%%%%%%%%%%%%%%%%%%%%%%%%%%%%%%%%%%%%%%%%%%%%%%%%%%%%%%%%%%%%%%
\section{MSE Analysis}
\label{sec:MSEanalysis}
In this section, we will analyze the MSE performance of the proposed GSE. Since the proposed GSE has similar forms in the parameter estimation scenario and the interference suppression scenario, we carried out the following derivations based on the parameter estimation scenario. Firstly, we will prove that the minimum MSE obtained by the GSE schemes will always be smaller or equal to the MSE that can be achieved by minimum variance unbiased estimator (MVUE) such as the LS estimator. Then the MSE lower bounds of the GSE schemes will be derived. In addition, we will prove that when the numbers of groups is larger than or equal to two, the MSE lower bound will always be lower than the biased estimator with only one shrinkage factor (when the number of groups $S$ equals one).

\subsection{MMSE Comparison}
Assuming AWGN with zero mean and variance $\sigma_{n}^2$, the LS estimator is a minimum variance unbiased estimator.
The MSE for the LS estimator is
%%%%%%%%%%%%%%%%%%%%%%%%%%%%%%%%%%%%%%%%%%%%%%%%%%%%%%%%%%%%%%%%%%%%%%%%%%%%%%%%%%%%%%%%%%%%%%%%%%%%%%%%%%%%%%%%%%%
\begin{equation}
{\mathbb E}\{\|\bm h - \hat{\bm h}_{\rm LS}\|^2\}={\mathbb E}\{(\hat{\bm
h}_{\rm LS}-\bm h)^{H}(\hat{\bm h}_{\rm LS}-\bm h)\}=var(\bm
h,\hat{\bm h}_{\rm LS}).
\end{equation} Defining $v=var(\bm
h,\hat{\bm h}_{\rm LS})$, we have
%%%%%%%%%%%%%%%%%%%%%%%%%%%%%%%%%%%%%%%%%%%%%%%%%%%%%%%%%%%%%%%%%%%%%%%%%%%%%%%%%%%%%%%%%%%%%%%%%%%%%%%%%%%%%%%%%%%
\begin{equation}
%\begin{split}
v =tr\{\sigma_{n}^2(\bm X^{H}\bm
X)^{-1}\}
%&= {\mathbb E}\{(\hat{\bm h}_{\rm LS}-\bm h)^{H}(\hat{\bm h}_{\rm
%LS}-\bm h)\}
%%={\mathbb E}\{\big((\bm X^{H}\bm X)^{-1}\bm X^{H}\bm
%%n\big)^{H}\big((\bm X^{H}\bm X)^{-1}\bm X^{H}\bm
%%n\big)\}\\
%=tr\{{\mathbb E}\{\big((\bm X^{H}\bm X)^{-1}\bm X^{H}\bm
%n\big)^{H}\big((\bm X^{H}\bm X)^{-1}\bm X^{H}\bm
%n\big)\}\}\\
%%&={\rm E}[tr\{\big((\mathbf X^{H}\mathbf X)^{-1}\mathbf X^{H}\mathbf
%%n\big)^{H}\big((\mathbf X^{H}\mathbf X)^{-1}\mathbf X^{H}\mathbf
%%n\big)\}]\\
%&={\mathbb E}\{tr\{\big((\bm X^{H}\bm X)^{-1}\bm X^{H}\bm
%n\big)\big((\bm X^{H}\bm X)^{-1}\bm X^{H}\bm
%n\big)^{H}\}\}=tr\{\sigma_{n}^2(\bm X^{H}\bm
%X)^{-1}\}\\
%\end{split}
\label{equ:v_6}
\end{equation}

It should be noted that we can also express the LS estimator as $\hat{\bm h}_{\rm LS}=\bm h+\tilde{\bm n}$, hence, we have that the variance of the elements in the equivalent noise $\tilde{\bm n}$ is $\tilde {\sigma}_{n}^{2}=v/L$.

Recall that the target of the biased estimation that is to reduce the MSE introduced by $\hat{\bm h}_{\rm LS}$. The objective is to obtain a biased estimator that results in
%%%%%%%%%%%%%%%%%%%%%%%%%%%%%%%%%%%%%%%%%%%%%%%%%%%%%%%%%%%%%%%%%%%%%%%%%%%%%%%%%%%%%%%%%%%%%%%%%%%%%%%%%%%%%%%%%%%
\begin{equation}
{\mathbb E}\{\|\bm h - \hat{\bm h}_{\rm b}\|^2\}\leq{\mathbb E}\{\|\bm h - \hat{\bm h}_{\rm LS}\|^2\}.
\label{equ:targetofbiased}
\end{equation}

%Now, the minimum MSE that corresponds to the optimal solution $\boldsymbol \alpha_{\rm opt}$ will be detailed.
Recall the equations \eqref{equ:varpartofMSE} and \eqref{equ:biasepartofMSE}, the objective becomes
%%%%%%%%%%%%%%%%%%%%%%%%%%%%%%%%%%%%%%%%%%%%%%%%%%%%%%%%%%%%%%%%%%%%%%%%%%%%%%%%%%%%%%%%%%%%%%%%%%%%%%%%%%%%%%%%%%%
\begin{equation}
\tilde {\sigma}^{2}({\bf 1}_{S}+\boldsymbol \alpha)^{H}({\bf 1}_{S}+\boldsymbol \alpha)+{\boldsymbol \alpha}^{H}\bm H^{H}\bm H{\boldsymbol \alpha}\leq v.
\label{equ:targetofbiased_2}
\end{equation}Since $\tilde {\sigma}_{n}^{2}=v/L$, we have $\tilde {\sigma}^{2}=(L/S)\tilde {\sigma}_{n}^{2}=v/S$.
%Note that
%%%%%%%%%%%%%%%%%%%%%%%%%%%%%%%%%%%%%%%%%%%%%%%%%%%%%%%%%%%%%%%%%%%%%%%%%%%%%%%%%%%%%%%%%%%%%%%%%%%%%%%%%%%%%%%%%%%%
%\begin{equation}
%\bm H^{H}\bm H=\begin{bmatrix}
% \sum^{L/S}_{i=1} |h(i)|^2        & 0                                 & \dots    & 0\\
% 0                                & \sum^{2L/S}_{i=L/S+1} |h(i)|^2    & \dots    &\vdots\\
% \vdots                           & \vdots                            & \dots    &\vdots\\
% \vdots                           & \vdots                            & \cdots   &\vdots\\
%  0                                & \dots                            & 0        & \sum^{L}_{i=(S-1)L/S+1} |h(i)|^2\\
%\end{bmatrix}.
%\end{equation} To simplify the following derivation, we define scalar terms $h_{\Sigma,s}=\sum^{sL/S}_{i=(s-1)L/S+1} |h(i)|^2$.
In appendix \ref{app:proofs1}, we prove that this objective is always fulfilled with the optimal solution $\boldsymbol \alpha_{\rm opt}$ as shown in \eqref{equ:aopt}.

\subsection{MSE Lower Bound and the Effect of the Number of Groups}
It should be noted that a lower bound of the MSE performance of the proposed GSE schemes that corresponds to the optimal ${\boldsymbol \alpha}$ can be obtained as
%%%%%%%%%%%%%%%%%%%%%%%%%%%%%%%%%%%%%%%%%%%%%%%%%%%%%%%%%%%%%%%%%%%%%%%%%%%%%%%%%%%%%%%%%%%%%%%%%%%%%%%%%%%%%%%%%%%
\begin{equation}
\begin{split}
{\mathbb E}\{\|\bm h - \hat{\bm h}_{\rm b}\|^2\}&\ge (S+\sum^{S}_{s=1}\alpha_{\rm opt,s})\frac{v}{S}=v-\frac{\sum^{S}_{s=1}(\frac{v^2}{v+h_{\Sigma,s}S})}{S}\\
&=v-\sum^{S}_{s=1}\frac{v^2}{Sv+h_{\Sigma,s}S^2}.
\end{split}
\label{equ:mselowerbound}
\end{equation}Since the second term on the right hand-side is non-negative, it can be concluded that the MSE lower bound will always be smaller than or equal to the unbiased Cram$\acute{e}$r-Rao Lower Bound (CRLB), which is expressed as $v$ in this equation. Note that this expression can be considered as the relationship between the MSE lower bound of the GSE and the unbiased CRLB. In the case where the number of groups equals 1, we have $S=1$, $\tilde {\sigma}^{2}=v$ and $\alpha_{\rm opt}=-v/(v+\|\bm h\|^2)$. The lower bound becomes
%%%%%%%%%%%%%%%%%%%%%%%%%%%%%%%%%%%%%%%%%%%%%%%%%%%%%%%%%%%%%%%%%%%%%%%%%%%%%%%%%%%%%%%%%%%%%%%%%%%%%%%%%%%%%%%%%%%
\begin{equation}
{\mathbb E}\{\|\bm h - \hat{\bm h}_{\rm b}\|^2\}\ge \frac{1}{1+v/\|\bm h\|^2}v.
\end{equation} In this case, the one-group GSE scheme converges to our previously proposed shrinkage factor biased estimator \cite{myEW2011}, \cite{myICASSP2011}.

For the proposed GSE schemes, we prove the following statements:\\
$Statement$ 1: The MSE lower bound as shown in \eqref{equ:mselowerbound} with $S>1$ will always be lower than or equal to the lower bound for $S=1$. This statement indicates that our proposed GSE outperforms the biased estimator with only one shrinkage factor. The proof is detailed in appendix \ref{app:proofs}.\\
$Statement$ 2: The lowest MSE lower bound as shown in \eqref{equ:mselowerbound} can be obtained in $S=L$ case, where $L$ is the length of the parameter vector to be estimated. This statement indicates that the optimal performance can be obtained with the largest possible group number. The proof is detailed in Appendix \ref{app:proofs}.

The performance of the algorithm depends on the number of groups and the
scenario. If some a priori knowledge of the parameter vector to be
estimated is available, then a possible extension of the GSE is to develop
a method to determine the size of the groups. For example, the knowledge
of the expected value of the number of clusters of a UWB channel might
enable a more attractive tradeoff between the performance and the
complexity. Moreover, the increase in the number of groups can improve the
MSE performance. However, this comes with diminishing returns and an
increase in the computational complexity.

%%%%%%%%%%%%%%%%%%%%%%%%%%%%%%%%%%%%%%%%%%%%%%%%%%%%%%%%%%%%%%%%%%%%%%%%%%%%%%%%%%%%%%%%%%%%%%%%%%%%%%%%%%%%%%%%%%%
\section{Simulations}
\label{sec:simulations}
\begin{figure}[htb]
\begin{minipage}[h]{1.0\linewidth}
  \centering
  \centerline{\epsfig{figure=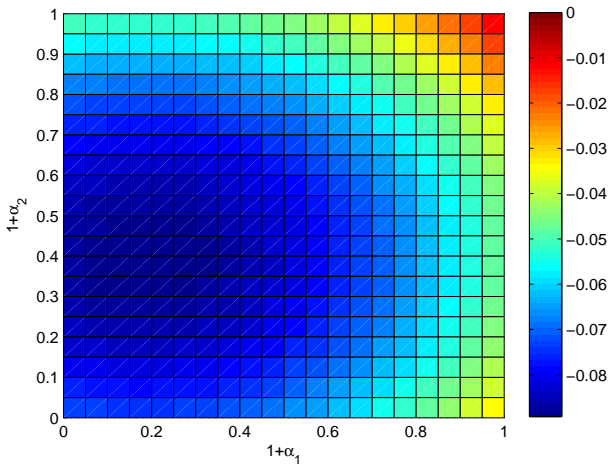,scale=0.6}}
\end{minipage}
\caption{Surface of the MSE difference in a scenario with $S=2$ (SCE). The MSE difference is defined as $\delta {\rm MSE}=\|\bm h - \hat{\bm h}_{\rm b}\|^2-\|\bm h - \hat{\bm h}_{\rm RLS}\|^2$.}
\label{fig:MSEsurface}
\end{figure}

\begin{figure}[htb]
\begin{minipage}[h]{1.0\linewidth}
  \centering
  \centerline{\epsfig{figure=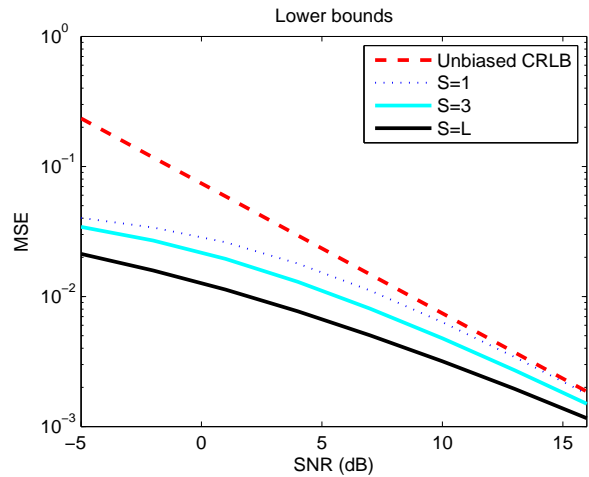,scale=0.7}}
\end{minipage}
\caption{MSE lower bounds for the proposed GSE with different numbers of groups (SCE).}
\label{fig:bounds}
\end{figure}

\begin{figure}[htb]
\begin{minipage}[h]{1.0\linewidth}
  \centering
  \centerline{\epsfig{figure=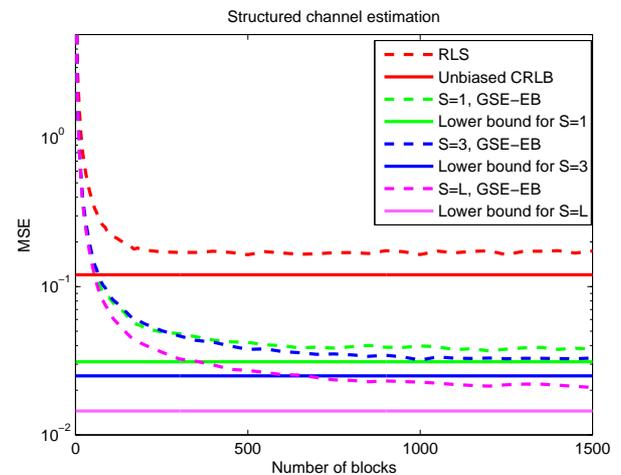,scale=0.6}}
\end{minipage}
\caption{MSE performance with different numbers of groups (SCE). Parameters used: RLS: $\lambda=0.998$, $\delta=10$. Proposed GSE EB: $\mu=0.075$.}
\label{fig:MSEvsS}
\end{figure}

\begin{figure}[htb]
\begin{minipage}[h]{1.0\linewidth}
  \centering
  \centerline{\epsfig{figure=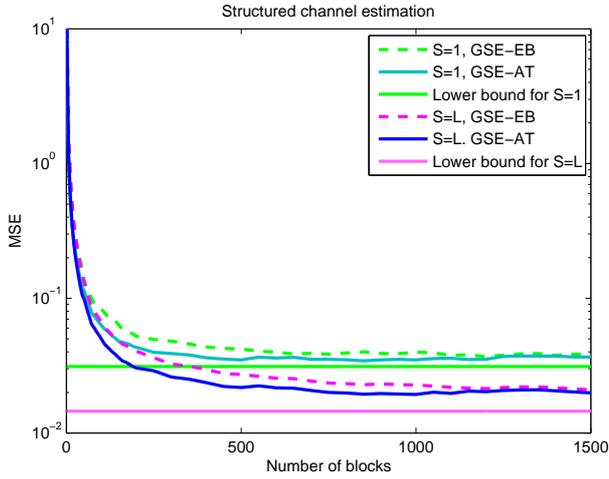,scale=0.6}}
\end{minipage}
\caption{MSE performance of the proposed GSE-EB and GSE-AT algorithms with $S=1$ and $S=L$ (SCE). Parameters used: Proposed GSE-EB: $\mu=0.075$. Proposed GSE-AT: $\mu=0.075$, $\mu_{p}=0.05$.}
\label{fig:MSEEBAT}
\end{figure}

\begin{figure}[htb]
\begin{minipage}[b]{1.0\linewidth}
  \centering
%  \centerline{\epsfig{figure=da_msesnr_NMSE.eps,width=8.5cm}}
  \centerline{\epsfig{figure=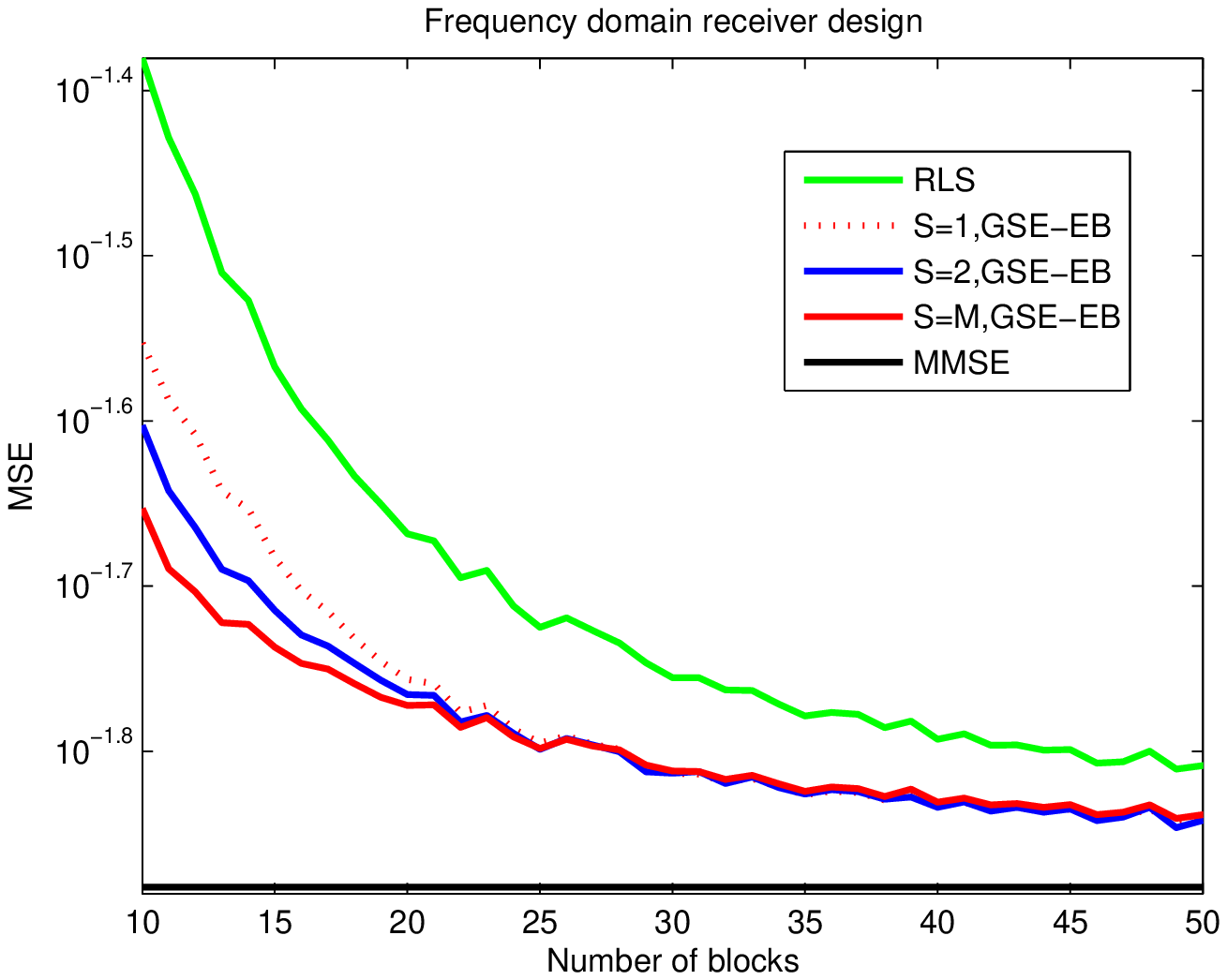,scale=0.6}}
\end{minipage}
\caption{Normalized MSE performance ($\|\bm b-\hat{\bm b}\|^2/\|\bm
b\|^2$) of the biased estimator with 5 users in 5 dB SNR.}
\label{fig:biasedceMSEsnrvsmse22}
\end{figure}

\begin{figure}[htb]
\begin{minipage}[b]{1.0\linewidth}
  \centering
  \centerline{\epsfig{figure=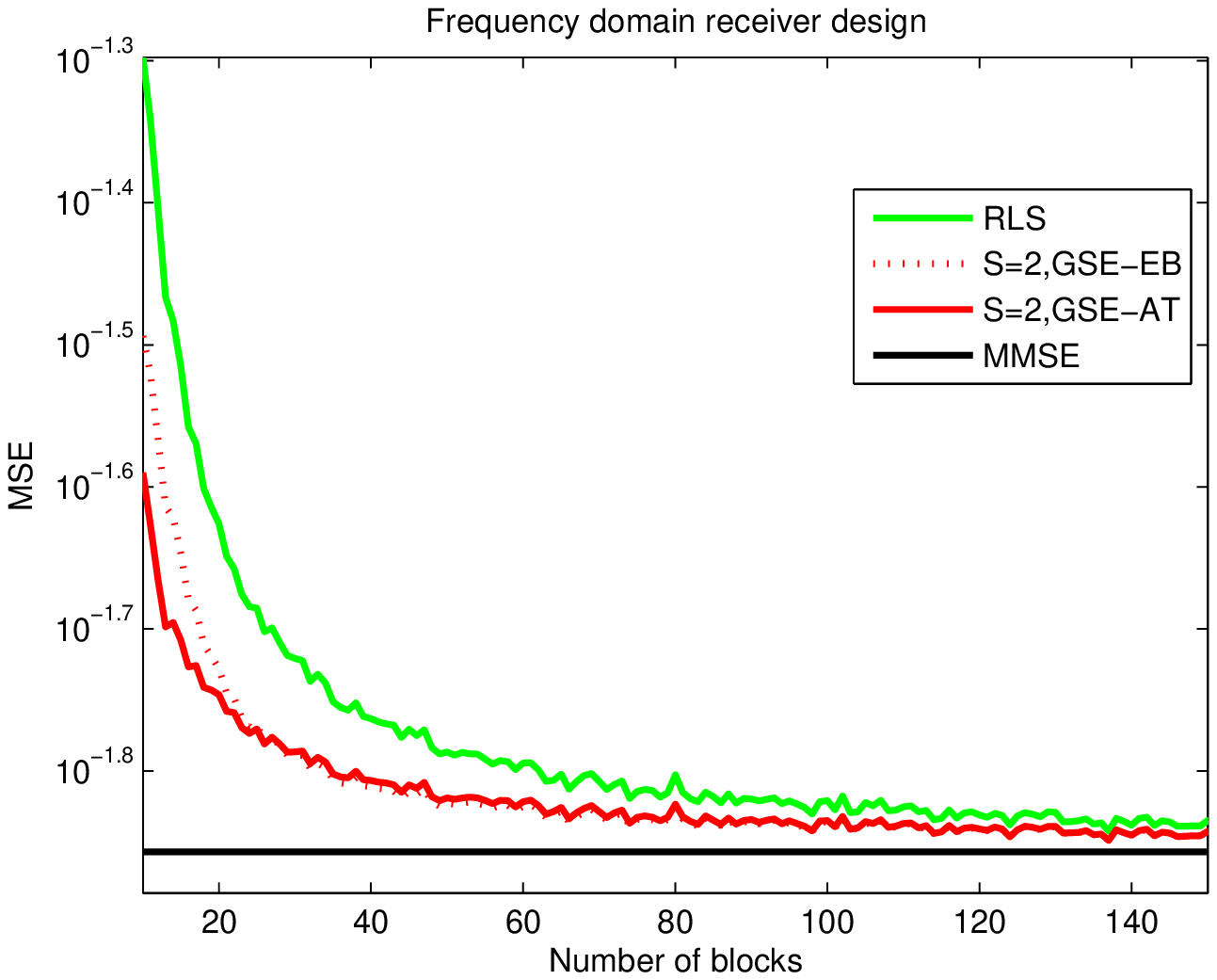,scale=0.6}}
\end{minipage}
\caption{Normalized MSE performance ($\|\bm b-\hat{\bm b}\|^2/\|\bm
b\|^2$) of the biased estimators with 5 users in 3 dB SNR.}
\label{fig:atebda}
\end{figure}

\begin{figure}[htb]
\begin{minipage}[b]{1.0\linewidth}
  \centering
  \centerline{\epsfig{figure=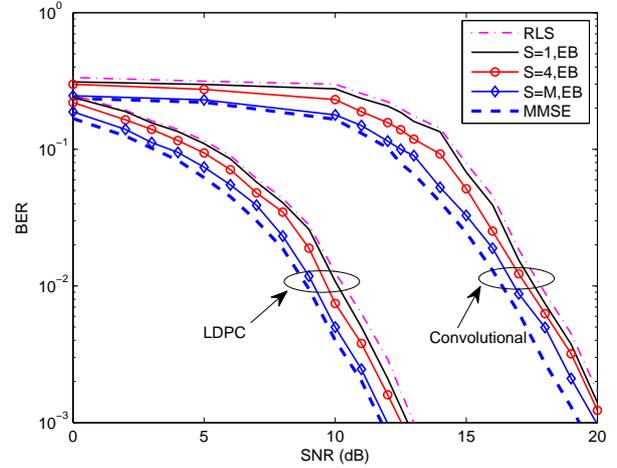,scale=0.6}}
\end{minipage}
\caption{Coded BER performance of the GSE schemes in interference suppression scenario with 5 users.}
\label{fig:berda}
\end{figure}

In this section, the proposed GSE estimators are employed in the SCE and in the design of the frequency domain receiver of a synchronous downlink block-by-block transmission binary phase shift keying (BPSK) DS-UWB system that are detailed in Section \ref{sec:systemmodelsce} and Section \ref{sec:systemmodelis}, respectively. Their MSE performances are compared with the conventional RLS adaptive algorithms. The pulse shape adopted is the root-raised-cosine (RRC) pulse with the pulse-width $T_{c}=0.375$ ns. The length of the data block is set to $N=32$ symbols. The Walsh spreading code with a spreading gain $N_{c}=8$ is generated for the simulations and we assume that the maximum number of active users is $7$. The channel has been simulated according to the standard IEEE 802.15.4a channel model for the NLOS indoor environment as shown in \cite{Molisch2006}. We assume that the channel is constant during the whole transmission and the time domain channel impulse response has $100$ taps. The CP guard interval has a length of $35$ chips, which has the equivalent length of $105$ samples and it is enough to eliminate the IBI. The uncoded data rate of the transmission is approximately $293$ $\rm Mbps$. For all the simulations, the adaptive receivers/estimators are initialized as null vectors. All the curves are obtained by averaging $200$ channel realizations. In Fig. \ref{fig:MSEsurface} - Fig. \ref{fig:MSEEBAT}, the performance of the proposed GSE in the parameter estimation scenario (SCE as an example) is shown. In this scenario, the GSE is employed to improve the MSE performance of the channel estimation. In Fig. \ref{fig:biasedceMSEsnrvsmse22} to Fig. \ref{fig:berda}, the performance of the proposed GSE in the interference suppression scenario (frequency domain receiver design as an example) is presented. In this scenario, the GSE schemes accelerate the convergence speed of the adaptive algorithm.

%%%%%%%%%%%%%%%%%%%%%%%%%%%%%%%%%%%%%%%%%%%%%%%%%%%%%%%%%%%%%%%%%%%%%
%%%%%%%%%%%%%%%%%%%%%%%%%%%%%%%%%%%%%%%%%%%%%%%%%%%%%%%%%%%%%%%%%%%%%
%%%%%%%%%%%%%%%%%%%%%%%%%%%%%%%%%%%%%%%%%%%%%%%%%%%%%%%%%%%%%%%%%%%%%
%%%%%%%%%%%%%%%%%%%%%%%%%%%%%%%%%%%%%%%%%%%%%%%%%%%%%%%%%%%%%%%%%%%%%
%%%%%%%%%%%%%%%%%%%%%%%%%%%%%%%%%%%%%%%%%%%%%%%%%%%%%%%%%%%%%%%%%%%%%
%%%%%%%%%%%%%%%%%%%%%%%%%%%%%%%%%%%%%%%%%%%%%%%%%%%%%%%%%%%%%%%%%%%%%
%%%%%%%%%%%%%%%%%%%%%%%%%%%%%%%%%%%%%%%%%%%%%%%%%%%%%%%%%%%%%%%%%%%%%
%%%%%%%%%%%%%%%%%%%%%%%%%%%%%%%%%%%%%%%%%%%%%%%%%%%%%%%%%%%%%%%%%%%%%
%%%%%%%%%%%%%%%%%%%%%%%%%%%%%%%%%%%%%%%%%%%%%%%%%%%%%%%%%%%%%%%%%%%%%%%%%%%%%%%%%%%%%%%%%%%%%%%%%%%%%%%%
%%%%%%%%%%%%%%%%%%%%%%%%%%%%%%%%%%%%%%%%%%%%%%%%%%%%%%%%%%%%%%%%%%%%%%%%%%%%%%%%%%%%%%%%%%%%%%%%%%%%%%%%
%%%%%%%%%%%%%%%%%%%%%%%%%%%%%%%%%%%%%%%%%%%%%%%%%%%%%%%%%%%%%%%%%%%%%%%%%%%%%%%%%%%%%%%%%%%%%%%%%%%%%%%%

First, we examine the MSE difference between the unbiased estimator and the proposed GSE as a function of the shrinkage factors for each group in a single user system with 0 dB SNR. In Fig. \ref{fig:MSEsurface}, the surface defined as $\delta {\rm MSE}=\|\bm h - \hat{\bm h}_{\rm b}\|^2-\|\bm h - \hat{\bm h}_{\rm RLS}\|^2$ in the range of the shrinkage factors between 0 to 1 is shown. Note that, when the bias equals zero, which corresponding to the point $[1,1]$ in the figure, the MSE difference equals zero. The optimal solution is located at $[1+\alpha_{1}, 1+\alpha_{2}]=[0.15,0.35]$. For this channel realization, the optimal solution that is obtained by our algorithm after transmitting 1000 data blocks is reported as $[1+\alpha_{1}, 1+\alpha_{2}]=[0.16,0.35]$, which is very close to the optimal solution shown in this figure.

In Fig. \ref{fig:bounds}, the MSE lower bounds for the proposed GSE schemes with different numbers of groups are shown as a function of SNR. The proposed biased estimators show a better performance than the conventional unbiased estimator. As we have proved, the best performance can be obtained in case of $S=L$. The proposed GSE schemes can maintain the MSE gain for low and medium SNR regimes. In high SNR scenarios with a small noise variance, the biased estimator converges to the unbiased estimator and the gain becomes smaller. Compared with the biased estimators with only one shrinkage factor (which is equivalent to the case $S=1$), the GSE schemes with a number of groups can maintain the MSE gain until higher SNR regime.

In the third experiment, we examine the proposed GSE schemes with different numbers of groups for the SCE in a single-user system with 0 dB SNR. In Fig. \ref{fig:MSEvsS}, the MSE performance of the channel estimators are compared as a function of the number of blocks transmitted. The RLS algorithm approaches the unbiased CRLB while the proposed biased estimators approach the lower bounds as given in \eqref{equ:mselowerbound}. The biased estimators converge faster than the RLS algorithm and the steady-state performance is also improved. Note that the additional complexity to employ the proposed biased estimation techniques increases linearly with the product of the number of groups and the length of the channel.

In Fig. \ref{fig:MSEEBAT}, the proposed GSE-EB and GSE-AT algorithms are compared in $S=1$ and $S=L$ scenarios in a single user system with 0 dB SNR. It can be found that the GSE-AT algorithm can provide a noticeable gain over the GSE-EB algorithm especially at the beginning stage of transmission. This is because the GSE-AT algorithm allows us to further adjust the diagonal matrix $\hat {\bm P}_{m}(i)$ for each time instant. At the beginning stage when the RLS algorithm does not very accurately estimate the channel, the GSE-AT algorithm can be used to improve the convergence rate.

In Fig. \ref{fig:biasedceMSEsnrvsmse22}, the performance of the proposed GSE is shown in the interference suppression scenario (frequency domain receiver design as an example) with a short training sequence. In this simulation, 50 training blocks are transmitted in a scenario with 5 users with 5 dB SNR. The proposed GSE-EB algorithm outperforms the RLS algorithm and the best performance is obtained by setting $S=M$. In Fig. \ref{fig:atebda}, we compare the GSE-EB and GSE-AT algorithms in 5-user communications and an SNR of 3 dB. In this experiment, 150 training blocks are transmitted. The GSE-AT algorithm can further accelerate the convergence rate of the GSE-EB algorithm. At the beginning stage, the GSE-AT algorithm introduces the best performance.

In Fig. \ref{fig:berda}, the bit error rate (BER) performance of the proposed GSE with different numbers of groups are shown in a scenario with 5 users. The coded BER performance is obtained by adopting a convolutional code and an LDPC code \cite{LDPC} designed according to the PEG approach \cite{LDPC}.  For the convolutional code, the constraint length is 5, the rate is 2/3 and the code polynomial is [7,5,5]. For the LDPC code, the rate is 1/2 and the code length is 200 bits. The maximum number of iterations is set to 20. In this experiment, 100 training blocks are transmitted followed by 400 data blocks. With the convolutional code, the proposed GSEs perform better than the RLS algorithm and the maximum gains are obtained in the medium SNR range from 12 dB to 17 dB. By employing the LDPC code, a BER of around $10^{-3}$ is achieved at 12 dB. The proposed GSEs with LDPC codes also outperform the RLS algorithm and the maximum gains are obtained for an SNR around 7 dB.

%

%%%%%%%%%%%%%%%%%%%%%%%%%%%%%%%%%%%%%%%%%%%%%%%%%%%%%%%%%%%%%%%%%%%%%%%%%%%%%%%%%%%%%%%%%%%%%%%%%%%%%%%%%%%%%%%%%%%
%%%%%%%%%%%%%%%%%%%%%%%%%%%%%%%%%%%%%%%%%%%%%%%%%%%%%%%%%%%%%%%%%%%%%%%%%%%%%%%%%%%%%%%%%%%%%%%%%%%%%%%%%%%%%%%%%%%
\section{Conclusions}
\label{sec:conclusion}

In this work, a novel biased estimation algorithm called group-based shrinkage estimator (GSE) is proposed, which divides the target parameter vector into a number of groups and calculates one shrinkage factor for each group. Adaptive algorithms are developed for the GSE scheme in the parameter estimation and the interference suppression scenarios. The incorporation of the proposed estimators has been considered in the frequency-domain of DS-UWB systems, where structured channel estimation and the receiver designs are considered as examples of the parameter estimation scenario and interference suppression scenario, respectively. An MSE analysis is presented that indicates the lower bound of the proposed GSE schemes. The relationship between the lower bound and the number of groups are also established. It has been proved that the GSE provides a better performance than the biased estimators with only one shrinkage factor. In addition, the lowest MSE lower bound can be obtained in the $S=L$ case. As for future research directions, the GSE scheme can be developed in different systems and scenarios. In addition, if we have some prior knowledge of the target parameter vector, we can then divide it into groups with different sizes and find more attractive tradeoffs between the computational complexity and the performance.

\appendices
%%%%%%%%%%%%%%%%%%%%%%%%%%%%%%%%%%%%%%%%%%%%%%%%%%%%%%%%%%%%%%%%%%%%%%%%%%%%%%%%%%%%%%%%%%%%%%%%%%%%%%%%
\section{MMSE Comparison}
\label{app:proofs1}
In order to check if the objective as shown in \eqref{equ:targetofbiased_2} is fulfilled with $\bm a_{\rm
opt}$, the equation \eqref{equ:aopt} is rearranged by taking the
expression of the diagonal matrix $\bm H^{H}\bm H\in \mathbb
C^{S\times S}$ into account. We have
%%%%%%%%%%%%%%%%%%%%%%%%%%%%%%%%%%%%%%%%%%%%%%%%%%%%%%%%%%%%%%%%%%%%%%%%%%%%%%%%%%%%%%%%%%%%%%%%%%%%%%%%%%%%%%%%%%%
\begin{equation}
\begin{split}
&\boldsymbol \alpha_{\rm opt} =[\alpha_{\rm opt,1},\dots,\alpha_{\rm opt,S}]^{T}= - \tilde {\sigma}^{2} \big(\tilde {\sigma}^{2}\bm I+\bm H^{H}\bm H\big)^{-1}{\bf 1}_{S}\\
%&=- \tilde {\sigma}^{2}[(\tilde {\sigma}^{2}+h_{\Sigma,1})^{-1},(\tilde {\sigma}^{2}+h_{\Sigma,2})^{-1},\dots,(\tilde {\sigma}^{2}+h_{\Sigma,S})^{-1}]^{T}.\\
&=-[(1+h_{\Sigma,1}/\tilde {\sigma}^{2})^{-1},(1+h_{\Sigma,2}/\tilde {\sigma}^{2})^{-1},\dots,(1+h_{\Sigma,S}/\tilde {\sigma}^{2})^{-1}]^{T}\\
\end{split}
\end{equation} where $h_{\Sigma,s}=\sum^{sL/S}_{i=(s-1)L/S+1} |h(i)|^2$. Hence, we have the following expressions
%%%%%%%%%%%%%%%%%%%%%%%%%%%%%%%%%%%%%%%%%%%%%%%%%%%%%%%%%%%%%%%%%%%%%%%%%%%%%%%%%%%%%%%%%%%%%%%%%%%%%%%%%%%%%%%%%%%
\begin{equation}
\tilde {\sigma}^{2}({\bf 1}_{S}+\boldsymbol \alpha)^{H}({\bf 1}_{S}+\boldsymbol \alpha)=\tilde {\sigma}^{2}\sum^{S}_{s=1}(1+\alpha_{\rm opt,s})^{2},
\label{equ:MSEinequ1}
\end{equation}
%%%%%%%%%%%%%%%%%%%%%%%%%%%%%%%%%%%%%%%%%%%%%%%%%%%%%%%%%%%%%%%%%%%%%%%%%%%%%%%%%%%%%%%%%%%%%%%%%%%%%%%%%%%%%%%%%%%
\begin{equation}
{\boldsymbol \alpha}^{H}\bm H^{H}\bm H{\boldsymbol \alpha}=\sum^{S}_{s=1}\alpha_{\rm opt,s}^{2}h_{\Sigma,s}.
\label{equ:MSEinequ2}
\end{equation} By substituting \eqref{equ:MSEinequ1} and \eqref{equ:MSEinequ2} into the MSE expression \eqref{equ:targetofbiased_2} and bearing in mind that $S\tilde {\sigma}^{2}=v$, the inequality that we want to prove becomes
%%%%%%%%%%%%%%%%%%%%%%%%%%%%%%%%%%%%%%%%%%%%%%%%%%%%%%%%%%%%%%%%%%%%%%%%%%%%%%%%%%%%%%%%%%%%%%%%%%%%%%%%%%%%%%%%%%%
\begin{equation}
\tilde {\sigma}^{2}\sum^{S}_{s=1}(1+\alpha_{\rm opt,s})^{2}+\sum^{S}_{s=1}\alpha_{\rm opt,s}^{2}h_{\Sigma,s}-S\tilde {\sigma}^{2} \leq 0.
\label{equ:MSEinequ3}
\end{equation} Note that the left hand-side of \eqref{equ:MSEinequ3} can be expressed as
%%%%%%%%%%%%%%%%%%%%%%%%%%%%%%%%%%%%%%%%%%%%%%%%%%%%%%%%%%%%%%%%%%%%%%%%%%%%%%%%%%%%%%%%%%%%%%%%%%%%%%%%%%%%%%%%%%%
\begin{equation}
\begin{split}
&\sum^{S}_{s=1}\alpha_{\rm opt,s}^{2}+\sum^{S}_{s=1}2\alpha_{\rm opt,s}+\sum^{S}_{s=1}\alpha_{\rm opt,s}^{2}h_{\Sigma,s}/\tilde {\sigma}^{2}\\
&=\sum^{S}_{s=1}\alpha_{\rm opt,s}^{2}(1+h_{\Sigma,s}/\tilde {\sigma}^{2})+\sum^{S}_{s=1}2\alpha_{\rm opt,s}.
\end{split}
\end{equation}%Note that the left-hand side terms can be expressed as
%%%%%%%%%%%%%%%%%%%%%%%%%%%%%%%%%%%%%%%%%%%%%%%%%%%%%%%%%%%%%%%%%%%%%%%%%%%%%%%%%%%%%%%%%%%%%%%%%%%%%%%%%%%%%%%%%%%%
%\begin{equation}
%\sum^{S}_{s=1}\alpha_{\rm opt,s}^{2}(1+h_{\Sigma,s}/\tilde {\sigma}^{2})+\sum^{S}_{s=1}2\alpha_{\rm opt,s}\leq 0.
%\end{equation}
By recalling that $\alpha_{\rm opt,s}=-(1+h_{\Sigma,s}/\tilde {\sigma}^{2})^{-1}$ is always a non-positive scalar value, the left-hand side of the equation becomes
%%%%%%%%%%%%%%%%%%%%%%%%%%%%%%%%%%%%%%%%%%%%%%%%%%%%%%%%%%%%%%%%%%%%%%%%%%%%%%%%%%%%%%%%%%%%%%%%%%%%%%%%%%%%%%%%%%%
\begin{equation}
\sum^{S}_{s=1}(1+h_{\Sigma,s}/\tilde {\sigma}^{2})^{-1}+\sum^{S}_{s=1}2\alpha_{\rm opt,s}=\sum^{S}_{s=1}\alpha_{\rm opt,s}.
\label{equ:mselowerboundbf}
\end{equation}
As $\alpha_{\rm opt,s}$ are always non-positive, the summation on the right hand-side of \eqref{equ:mselowerboundbf} will always be
smaller or equal to zero, which completes the proof.

\section{Proofs of the statements}
\label{app:proofs}
First, we want to prove that the lower bound of MSE as shown in \eqref{equ:mselowerbound} with $S>1$ will always be lower than or equal to the $S=1$ case. Based on the equation \eqref{equ:mselowerbound}, we can focus on the following function
%%%%%%%%%%%%%%%%%%%%%%%%%%%%%%%%%%%%%%%%%%%%%%%%%%%%%%%%%%%%%%%%%%%%%%%%%%%%%%%%%%%%%%%%%%%%%%%%%%%%%%%%%%%%%%%%%%%
\begin{equation}
f(S)=\sum^{S}_{s=1}\frac{1}{Sv+h_{\Sigma,s}S^2}=\frac{1}{S^2}\sum^{S}_{s=1}\frac{1}{(v/S)+h_{\Sigma,s}},
\end{equation} where $h_{\Sigma,s}=\sum^{sL/S}_{i=(s-1)L/S+1} |h(i)|^2$ The task now is equivalent to proving that $f(S)\ge f(1)$ for all the possible values of $S$. Note that $\sum^{S}_{s=1}h_{\Sigma,s}=\|\bm h\|^2$ and
%%%%%%%%%%%%%%%%%%%%%%%%%%%%%%%%%%%%%%%%%%%%%%%%%%%%%%%%%%%%%%%%%%%%%%%%%%%%%%%%%%%%%%%%%%%%%%%%%%%%%%%%%%%%%%%%%%%
\begin{equation}
f(1)=\frac{1}{v+\|\bm h\|^2}=\frac{1}{v+\sum^{S}_{s=1}h_{\Sigma,s}}=\frac{1}{\sum^{S}_{s=1}(v/S+h_{\Sigma,s})}.
\end{equation}Hence, the relation becomes
%%%%%%%%%%%%%%%%%%%%%%%%%%%%%%%%%%%%%%%%%%%%%%%%%%%%%%%%%%%%%%%%%%%%%%%%%%%%%%%%%%%%%%%%%%%%%%%%%%%%%%%%%%%%%%%%%%%
\begin{equation}
\sum^{S}_{s=1}\frac{1}{(v/S)+h_{\Sigma,s}}\ge \frac{S^2}{\sum^{S}_{s=1}(v/S+h_{\Sigma,s})}.
\end{equation} Actually, we can express this problem as the following mathematical problem:
%%%%%%%%%%%%%%%%%%%%%%%%%%%%%%%%%%%%%%%%%%%%%%%%%%%%%%%%%%%%%%%%%%%%%%%%%%%%%%%%%%%%%%%%%%%%%%%%%%%%%%%%%%%%%%%%%%%
\begin{equation}
\frac{1}{a_{1}}+\frac{1}{a_{2}}+\dots+\frac{1}{a_{S}}\ge \frac{S^2}{\sum^{S}_{s=1}a_{s}},
\end{equation}where $S$ is a positive integer and $a_{s}=(v/S)+h_{\Sigma,s}$ are all non-negative values.
This inequality can be proved by using the mathematical induction as follows:\\
For $S=1$, the left-hand side and the right-hand side are both equal to $\frac{1}{a_{1}}$.\\
For $S=2$, the left-hand side equals $\frac{1}{a_{1}}+\frac{1}{a_{2}}=\frac{a_{1}+a_{2}}{a_{1}a_{2}}$ and the right-hand side equals $\frac{4}{a_{1}+a_{2}}$. Since $(a_{1}+a_{2})^2-4a_{1}a_{2}=(a_{1}-a_{2})^2\ge 0$, the inequality holds.\\
Assuming the inequality holds for $S=n$, where $n\ge 2$, we have
%%%%%%%%%%%%%%%%%%%%%%%%%%%%%%%%%%%%%%%%%%%%%%%%%%%%%%%%%%%%%%%%%%%%%%%%%%%%%%%%%%%%%%%%%%%%%%%%%%%%%%%%%%%%%%%%%%%
\begin{equation}
\frac{1}{a_{1}}+\frac{1}{a_{2}}+\dots+\frac{1}{a_{n}}\ge \frac{n^2}{\sum^{n}_{s=1}a_{s}}.
\end{equation}
For $S=n+1$, we first consider the left-hand side as follows
%%%%%%%%%%%%%%%%%%%%%%%%%%%%%%%%%%%%%%%%%%%%%%%%%%%%%%%%%%%%%%%%%%%%%%%%%%%%%%%%%%%%%%%%%%%%%%%%%%%%%%%%%%%%%%%%%%%
\begin{equation}
\begin{split}
\frac{1}{a_{1}}+\frac{1}{a_{2}}+\dots+\frac{1}{a_{n}}+\frac{1}{a_{n+1}}&\ge \frac{n^2}{\sum^{n}_{s=1}a_{s}}+\frac{1}{a_{n+1}}\\
&=\frac{n^2 a_{n+1}+\sum^{n}_{s=1}a_{s}}{(\sum^{n}_{s=1}a_{s})\cdot a_{n+1}}.
\end{split}
\end{equation} and then, the right-hand side is given by
%%%%%%%%%%%%%%%%%%%%%%%%%%%%%%%%%%%%%%%%%%%%%%%%%%%%%%%%%%%%%%%%%%%%%%%%%%%%%%%%%%%%%%%%%%%%%%%%%%%%%%%%%%%%%%%%%%%
\begin{equation}
\frac{(n+1)^2}{\sum^{n}_{s=1}a_{s}+a_{n+1}}.
\end{equation} Because
%%%%%%%%%%%%%%%%%%%%%%%%%%%%%%%%%%%%%%%%%%%%%%%%%%%%%%%%%%%%%%%%%%%%%%%%%%%%%%%%%%%%%%%%%%%%%%%%%%%%%%%%%%%%%%%%%%%
\begin{equation*}
\begin{split}
&\big(\sum^{n}_{s=1}a_{s}+a_{n+1}\big)\big(n^2 a_{n+1}+\sum^{n}_{s=1}a_{s}\big)-(n+1)^2\big(\sum^{n}_{s=1}a_{s}\big)\cdot a_{n+1}\\
&=\big(\sum^{n}_{s=1}a_{s}-n\cdot a_{n+1}\big)^2\ge 0,
\end{split}
\end{equation*} the inequality also holds for $S=n+1$. This completes the proof.

Now, we can prove the second statement which points out that the lowest MSE lower bound as shown in \eqref{equ:mselowerbound} can be obtained when the numbers of groups is equal to the length of the parameter vector to be estimated. Following the proof of of $Statement$ 1, we can express the problem that needs to be solved as: prove that $f(L)\ge f(S)$, for any possible values of number of groups $S$, and mathematically, we need to prove that
%%%%%%%%%%%%%%%%%%%%%%%%%%%%%%%%%%%%%%%%%%%%%%%%%%%%%%%%%%%%%%%%%%%%%%%%%%%%%%%%%%%%%%%%%%%%%%%%%%%%%%%%%%%%%%%%%%%
\begin{equation}
\frac{1}{L^2}\sum^{L}_{i=1}\frac{1}{(v/L)+|h(i)|^2}\ge \frac{1}{S^2}\sum^{S}_{s=1}\frac{1}{(v/S)+h_{\Sigma,s}},
\end{equation} holds for all the possible values of $S$. Since the parameter vector is divided into a number of $S$ groups,
this inequality holds if the following relationship is fulfilled:
%%%%%%%%%%%%%%%%%%%%%%%%%%%%%%%%%%%%%%%%%%%%%%%%%%%%%%%%%%%%%%%%%%%%%%%%%%%%%%%%%%%%%%%%%%%%%%%%%%%%%%%%%%%%%%%%%%%
\begin{equation}
\begin{split}
&\frac{1}{L^2}\sum^{sL/S}_{i=(s-1)L/S+1}\frac{1}{(v/L)+|h(i)|^2}\ge \frac{1}{S^2}\frac{1}{(v/S)+h_{\Sigma,s}},\\
&\hspace{1em}with \hspace{0.5em}s=1,2,\dots,S
\end{split}
\label{equ:inequalitySL}
\end{equation}
Actually, the $S=L$ case can be considered as the division of each group (with length $L/S$) into a number of $L/S$ length-one sub-groups, and the inequality of \eqref{equ:inequalitySL} always holds for each group because of $Statement$ 1. This completes the proof.

%%%%%%%%%%%%%%%%%%%%%%%%%%%%%%%%%%%%%%%%%%%%%%%%%%%%%%%%%%%%%%%%%%%%%%%%%%%%%%%%%%%%%%%%%%%%%%%%%%%%%%%%

%
\begin{IEEEbiography}{Sheng Li}
(S'08 - M'11) received his Bachelor Degree in Zhejiang University of Technology in China in 2006 and the M.Sc. degree in communications engineering and Ph.D. degree in electronics engineering, both from the University of York, U.K. in 2007 and 2010, respectively. In 2010, he was awarded the K. M. Stott prize for excellence in scientific research. During November 2010 to October 2011, he has carried out a postdoctoral research in the Ilmenau University of Technology, Germany. Since April 2012, he has been with Zhejiang University of Technology, China, where he is currently a lecturer.
\end{IEEEbiography}

\begin{IEEEbiography}{Rodrigo C. de Lamare }
(S'99 - M'04 - SM'10) received the Diploma in electronic engineering from the Federal University of Rio de Janeiro (UFRJ) in 1998 and the M.Sc. and PhD degrees, both in electrical engineering, from the Pontifical Catholic University of Rio de Janeiro (PUC-Rio) in 2001 and 2004, respectively. Since January 2006, he has been with the Communications Research Group, Department of Electronics, University of York, where he is currently a Reader. Since April 2011, he has also been a Professor with PUC-RIO. His research interests lie in communications and signal processing, areas in which he has published nearly 300 papers in refereed journals and conferences. Dr. de Lamare serves as associate editor for the EURASIP Journal on Wireless Communications and Networking. He is a Senior Member of the IEEE has served as the General Chair of the 7th IEEE International Symposium on Wireless Communications Systems (ISWCS), held in York, UK in September 2010, and will serve as the Technical Programme Chair of ISWCS 2013 in Ilmenau, Germany.
\end{IEEEbiography}

\begin{IEEEbiography}{Martin Haardt}
(S'90 - M'98 - SM'99) has been a Full Professor in the Department of Electrical Engineering and Information Technology and Head of the Communications Research Laboratory at Ilmenau University of Technology, Germany, since 2001. Since 2012, he has also served as an Honorary Visiting Professor in the Department of Electronics at the University of York, UK.\\
After studying electrical engineering at the Ruhr-University Bochum, Germany, and at Purdue University, USA, he received his Diplom-Ingenieur (M.S.) degree from the Ruhr-University Bochum in 1991 and his Doktor-Ingenieur (Ph.D.) degree from Munich University of Technology in 1996.\\
In 1997 he joint Siemens Mobile Networks in Munich, Germany, where he was responsible for strategic research for third generation mobile radio systems. From 1998 to 2001 he was the Director for International Projects and University Cooperations in the mobile infrastructure business of Siemens in Munich, where his work focused on mobile communications beyond the third generation. During his time at Siemens, he also taught in the international Master of Science in Communications Engineering program at Munich University of Technology.\\
Martin Haardt has received the 2009 Best Paper Award from the IEEE Signal Processing Society, the Vodafone (formerly Mannesmann Mobilfunk) Innovations-Award for outstanding research in mobile communications, the ITG best paper award from the Association of Electrical Engineering, Electronics, and Information Technology (VDE), and the Rohde and Schwarz Outstanding Dissertation Award. In the fall of 2006 and the fall of 2007 he was a visiting professor at the University of Nice in Sophia-Antipolis, France, and at the University of York, UK, respectively. His research interests include wireless communications, array signal processing, high-resolution parameter estimation, as well as numerical linear and multi-linear algebra.\\
Prof. Haardt has served as an Associate Editor for the IEEE Transactions on Signal Processing (2002-2006 and since 2011), the IEEE Signal Processing Letters (2006-2010), the Research Letters in Signal Processing (2007-2009), the Hindawi Journal of Electrical and Computer Engineering (since 2009), the EURASIP Signal Processing Journal (since 2011), and as a guest editor for the EURASIP Journal on Wireless Communications and Networking. He has also served as an elected member of the Sensor Array and Multichannel (SAM) technical committee of the IEEE Signal Processing Society (since 2011), as the technical co-chair of the IEEE International Symposiums on Personal Indoor and Mobile Radio Communications (PIMRC) 2005 in Berlin, Germany, as the technical program chair of the IEEE International Symposium on Wireless Communication Systems (ISWCS) 2010 in York, UK, as the general chair of ISWCS 2013 in Ilmenau, Germany, and as the general-co chair of the 5-th IEEE International Workshop on Computational Advances in Multi-Sensor Adaptive Processing (CAMSAP) 2013 in Saint Martin, French Caribbean.
\end{IEEEbiography}

\end{document}